\documentclass[sigconf]{acmart}

\usepackage{geometry}
\usepackage{graphicx}
\usepackage{subfig}
\usepackage{caption}
\usepackage{hyperref}
\usepackage{amsmath}
\usepackage{multirow}
\usepackage{xspace}
\usepackage{threeparttable}

\usepackage{wasysym}

\let\oldhat\hat
\renewcommand{\vec}[1]{\mathbf{#1}}
\renewcommand{\hat}[1]{\oldhat{\mathbf{#1}}}

\newcommand{\ie}{\emph{i.e., }}
\newcommand{\eg}{\emph{e.g., }}

\newcommand{\etc}{\emph{etc}}
\newcommand{\wrt}{\emph{w.r.t. }}

\newcommand{\aka}{\emph{aka. }}

\AtBeginDocument{%
  \providecommand\BibTeX{{%
    \normalfont B\kern-0.5em{\scshape i\kern-0.25em b}\kern-0.8em\TeX}}}

\copyrightyear{2022}
\acmYear{2022}
\setcopyright{acmcopyright}
\acmConference[SIGIR '22] {Proceedings of the 45th International ACM SIGIR Conference on Research and Development in Information Retrieval}{July 11--15, 2022}{Madrid, Spain.}
\acmBooktitle {Proceedings of the 45th International ACM SIGIR Conference on Research and Development in Information Retrieval (SIGIR '22), July 11--15, 2022, Madrid, Spain}
\acmPrice{15.00}
\acmISBN{978-1-4503-8732-3/22/07}
\acmDOI{10.1145/3477495.3532025}

\begin{document}
\fancyhead{}

\title{Multi-level Cross-view Contrastive Learning for Knowledge-aware Recommender System}

\author{Ding Zou\textsuperscript{1,6},
Wei Wei\textsuperscript{1,6 $\XBox$},
Xian-Ling Mao\textsuperscript{2},
Ziyang Wang\textsuperscript{1,6},
Minghui Qiu\textsuperscript{3},
Feida Zhu\textsuperscript{4},
Xin Cao\textsuperscript{5}
}
\thanks{$\XBox$: Corresponding Author}
\affiliation{\textsuperscript{1} Cognitive Computing and Intelligent Information Processing (CCIIP) Laboratory, School of Computer Science and Technology, Huazhong University of Science and Technology
\country{China}}
\affiliation{\textsuperscript{2} School of Computer Science and Technology, Beijing Institute of Technology
\country{China}}
\affiliation{\textsuperscript{3} Alibaba Group
\country{China}}
\affiliation{\textsuperscript{4} Singapore Management University
\country{Singapore}}
\affiliation{\textsuperscript{5} School of Computer Science and Engineering, The University of New South Wales
\country{Australia}}
\affiliation{\textsuperscript{6} Joint Laboratory of HUST and Pingan Property \& Casualty Research (HPL)
\country{China}}
\affiliation{\textsuperscript{1} \{m202173662, weiw, ziyang1997 \}@hust.edu.cn
\quad\textsuperscript{2} maoxl@bit.edu.cn\quad \textsuperscript{3} minghuiqiu@gmail.com\quad\\ \textsuperscript{4} fdzhu@smu.edu.sg \quad\textsuperscript{5} xin.cao@unsw.edu.au
\country{}}

\begin{abstract}
Knowledge graph (KG) plays an increasingly important role in recommender systems.
Recently, graph neural networks (GNNs) based model has gradually become the theme of knowledge-aware recommendation (KGR).
However, there is a natural deficiency for GNN-based KGR models, that is, the sparse supervised signal problem, which may make their actual performance drop to some extent.
Inspired by the recent success of contrastive learning in mining supervised signals from data itself, in this paper, we focus on exploring the contrastive learning in KG-aware recommendation and propose a novel multi-level cross-view contrastive learning mechanism, named MCCLK.
Different from traditional contrastive learning methods which generate two graph views by uniform data augmentation schemes such as corruption or dropping, we comprehensively consider three different graph views for KG-aware recommendation, including global-level structural view, local-level collaborative and semantic views.
Specifically, we consider the user-item graph as a collaborative view, the item-entity graph as a semantic view, and the user-item-entity graph as a structural view.
MCCLK hence performs contrastive learning across three views on both local and global levels, mining comprehensive graph feature and structure information in a self-supervised manner.
Besides, in semantic view, a $k$-Nearest-Neighbor ($k$NN) item-item semantic graph construction module is proposed, to capture the important item-item semantic relation which is usually ignored by previous work.
Extensive experiments conducted on three benchmark datasets show the superior performance of our proposed method over the state-of-the-arts.
The implementations are available at: https://github.com/CCIIPLab/MCCLK.
\end{abstract}

\ccsdesc[500]{Information systems~Recommender systems}


\keywords{Graph Neural Network, Contrastive Learning, Knowledge Graph, Recommender System, Multi-view Graph Learning}

\maketitle

\section{Introduction}

Recommender system is crucial for users to discover items of interest in practice. Conventional recommendation approaches (\eg collaborative filtering (CF) \cite{he2017neural,lian2018xdeepfm,wang2019neural, liu2014exploiting}) rely on the availability of historical user behavior data (\eg user-item interactions~\cite{wang2020global,zhao22AAAI}) to capture collaborative signals for recommendation. However, they severely suffer from the cold-start problem, since they often treat each interaction as an independent instance while neglecting their relations, such as NFM \cite{he2017neural}, xDeepFM \cite{lian2018xdeepfm}. A widely-adopted solution is to incorporate various kinds of side information, such as knowledge graph (KG)~\cite{pan2021context}, which contains rich facts and connections about items, to learn high-quality user and item representations for recommendation (\aka knowledge-aware recommendation, KGR).

Indeed, there already exists much research effort \cite{wang2018dkn, zhang2016collaborative, zhang2018learning} devoted to KGR, the core of which is how to effectively leverage the graph of item side (\emph{heterogeneous}) information into the latent user/item representation learning. Most of early studies \cite{zhang2016collaborative, huang2018improving, wang2018shine,wang2018dkn} on KGR focus on employing different \emph{knowledge graph embedding} (KGE) models (\eg TransE \cite{bordes2013translating}, TransH \cite{wang2014knowledge}), to pre-train entity embeddings for item representation learning. However, these methods perform poorly, since they treat each item-entity relation independently for learning. Thus, the learning process is incapable of distilling sufficient collaborative signals for item representations.

Sequentially, many connection-based approaches are proposed to model multiple patterns of connections among user, item, and entity for recommendation, which can be further categorized into two classes, namely, \emph{path}-based \cite{hu2018leveraging, shi2018heterogeneous, wang2019explainable} and \emph{graph neural networks (GNN)} based \cite{hu2018leveraging, shi2018heterogeneous, wang2019explainable}.
The \emph{former} mainly focuses on enriching user-item interactions via capturing the long-range structure of KG, such as the selection of prominent paths over KG \cite{sun2018recurrent} \emph{or} representing the interactions with multi-hop paths from users to items \cite{hu2018leveraging, wang2019explainable}.
However, these methods heavily rely on manually designed meta-paths, and are thus hard to optimize in reality.
The \emph{later} is widely-adopted as an informative aggregation paradigm to integrate multi-hop neighbors into node representations \cite{sha2019attentive, wang2019knowledge, wang2019kgat, wang2021learning}, due to its powerful capability in effectively generating local permutation-invariant aggregation on the neighbors of a node for representation.
Despite effectiveness, current GNN-based models greatly suffer from sparse supervision signal problem, owing to the extreme sparsity of interactions \cite{bayer2017generic, wu2021self} and even terrible side effects, \eg degeneration problem \cite{gao2019representation}, \ie degenerating node embeddings distribution into a narrow cone, even leading to the indiscrimination of generated node representations.

However, alleviating the \emph{sparse supervision signal} problem faces a significant challenge, that is, the inadequacy of training labels, as labels are usually scarce in real recommendation applications.
Recently, contrastive learning, one of the classical Self-supervised learning (SSL) methods, is proposed to pave a way to enable training models without explicit labels \cite{liu2021self}, as its powerful capability in learning discriminative embeddings from unlabeled sample data, via maximizing the distance between negative samples while minimizing the distance between positive samples.
To this end, in this paper we mainly focus on designing an end-to-end knowledge-aware model within a contrastive learning paradigm, which requires us to sufficiently leverage the limited user-item interactions and additional KG facts (\eg item-entity affiliations) for recommendation.

Actually, it is still non-trivial to design a proper contrastive learning framework, for that characteristics of both contrastive learning and knowledge-aware recommendation are needed to be carefully considered for balance, which requires us to address the following fundamental issues \cite{wang2021self}:
(1) \emph{How to design a proper contrastive mechanism?}
Due to heterogeneity, the designed model is naturally required to simultaneously handle multiple types of nodes (\eg user, item, and entity) and relations (\eg user-item, item-entity and \etc).
(2) \emph{How to construct proper views for contrastive learning?}
A straightforward way is that, we can augment (or corrupt) the input user-item-entity graph as a graph view, and contrast it with the original one, analogous to \cite{chen2020simple, he2020momentum, lan2019albert}.
However, it is far from enough to solely consider global-level view (\ie user-item-entity graph) for KGR, because it is incapable of fully leveraging the rich collaborative information (\ie item-user-item co-occurrence) and semantic information (\ie item-entities-item co-occurrence).
Transparently, only utilizing one graph view (\eg user-item-entity graph) at a coarse-grained level makes it difficult in fully exploiting the rich collaborative and semantic information for recommendation.

\begin{figure}[t]
  \centering
  \includegraphics[width=\linewidth]{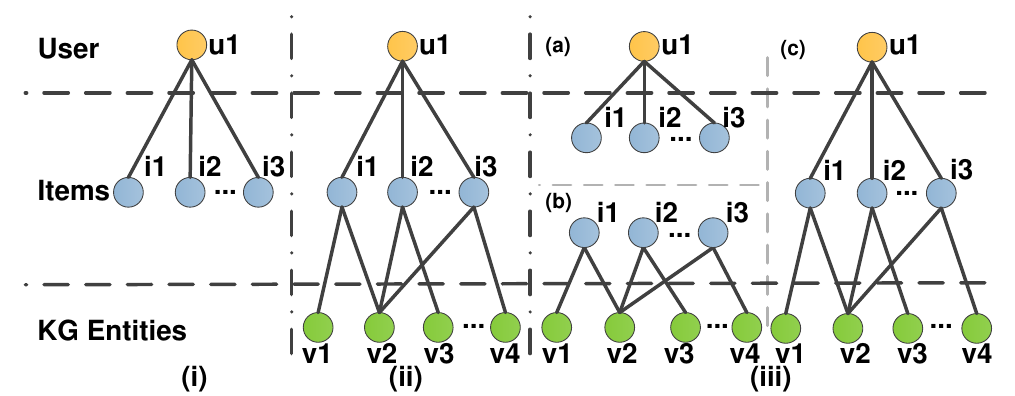}
  \caption{A toy example of our selected multi views. (i) Traditional CF-based recommendation learns from collaborative view. (ii) Previous KGR methods learn from the structural view. (iii) MCCLK learns from three selected views, including local-level collaborative view (a) and semantic view (b), global-level structural view (c).}
  \label{fig:toy}
  \vspace{-0.4cm}
\end{figure}

In this paper, we emphasize that the designed model should explore more graph views for learning in a more fine-grained manner.
Besides, since the considered distinct graph views may be in different levels, it's not feasible to simply contrast them at the same level, and thus a multi-level cross-view contrastive mechanism is inevitably important for the model designing.
Therefore, this paper proposes a novel model based on the self-supervised learning paradigm, named \textbf{M}ulti-level \textbf{C}ross-view \textbf{C}ontrastive \textbf{L}earning for \textbf{K}nowledge-aware Recommender System (MCCLK),
to fully leverage the rich collaborative and semantic information over KG and user-item interactions for KGR.
Specifically, we first comprehensively consider \textbf{three complementary graph views}. As shown in Figure \ref{fig:toy}, we consider the user-item graph as collaborative view and item-entity graph as semantic view, both of which are local-level views. Besides, to preserve complete structural information (\ie long-range user-item-entity connections), user-item-entity graph is considered as a structural view in global level. Then a novel \textbf{multi-level cross-view contrastive learning mechanism} is proposed to collaboratively supervise the three graph views, which performs local-level contrastive learning between collaborative view and semantic view, global-level contrastive learning between global-level and local-level views.
In particular, in the less explored semantic view, an effective $k$-Nearest-Neighbor ($k$NN) item-item semantic graph construction module is proposed, equipped with a relation-aware GNN, for explicitly considering item-item semantic similarity from knowledge information.
Moreover, adaptive GNN-based graph encoders are adopted for each graph view, stressing different parts of graph information according to the views' features.
Empirically, MCCLK outperforms the state-of-the-art models on three benchmark datasets.

\textbf{Our contributions} of this work can be summarized as follows:
\begin{itemize}
    \item \textbf{General Aspects:} We emphasize the importance of incorporating self-supervised learning into knowledge-aware recommendation, which takes node self-discrimination as a self-supervised task to offer auxiliary signal for graph representation learning.
    \item \textbf{Novel Methodologies:}
    We propose a novel model MCCLK, which builds a multi-level cross-view contrastive framework for knowledge-aware recommendation. MCCLK considers three views from user-item-entity graph, including global-level structural view, local-level collaborative and semantic views. MCCLK then performs local-level and global-level contrastive learning to enhance representation learning from multi-faced aspects. Moreover, in semantic view, a $k$NN item-item semantic graph construction module is proposed to explore item-item semantic relation.
    \item \textbf{Multifaceted Experiments:} We conduct extensive experiments on three benchmark datasets. The results demonstrate the advantages of our MCCLK in better representation learning, which shows the effectiveness of our multi-level cross-view contrastive learning framework and specially tailored graph aggregating mechanisms.
\end{itemize}

\section{Related Work}

\subsection{Knowledge-aware Recommendation}

\subsubsection{Embedding-based methods.} \label{embedding-based}
 Embedding-based methods \cite{cao2019unifying, wang2018dkn, zhang2016collaborative, huang2018improving, zhang2018learning, wang2018shine, wang2019multi} use knowledge graph embeddings(KGE) \cite{wang2014knowledge, bordes2013translating, lin2015learning} to preprocess a KG, then incorporate the learned entity embeddings and relation embeddings into the recommendation. Collaborative Knowledge base Embedding(CKE) \cite{zhang2016collaborative} combines CF module with structural, textual, and visual knowledge embeddings of items in a unified Bayesian framework. KTUP \cite{cao2019unifying} utilizes TransH \cite{wang2014knowledge} on user-item interactions and KG triplets to jointly learn user preference and perform KG completion. RippleNet \cite{wang2018ripplenet} explores users’ potential interests by propagating users’ historical clicked items along links (relations) in KG. Embedding-based methods show high flexibility in utilizing KG, but the KGE algorithms focus more on modeling rigorous semantic relatedness (\eg TransE \cite{bordes2013translating} assumes head + relation = tail), which are more suitable for link prediction rather than recommendation.

\subsubsection{Path-based methods.} \label{path-based}
Path-based methods \cite{yu2014personalized, hu2018leveraging, shi2018heterogeneous, wang2019explainable, yu2013collaborative, zhao2017meta} explore various patterns of connections among items in KG to provide additional guidance for the recommendation. For example, regarding KG as a Heterogeneous Information Network (HIN), Personalized Entity Recommendation (PER) \cite{yu2014personalized} and meta-graph based recommendation \cite{hu2018leveraging} extract the meta-path/meta-graph latent features and exploit the connectivity between users and items along different types of relation paths/graphs. KPRN \cite{wang2019explainable} further automatically extracts paths between users and items, and utilizes RNNs to model these paths. Path-based methods make use of KG in a more natural way, but they rely heavily on manually designed meta paths which can be hard to tune in reality. In addition, defining effective meta-paths requires domain knowledge, which is usually labor-intensive especially for complicated knowledge graphs.

\subsubsection{GNN-based methods.}
GNN-based methods \cite{yu2014personalized, hu2018leveraging, wang2019explainable, yu2013collaborative, zhao2017meta} are founded on the information aggregation mechanism of graph neural networks (GNNs) \cite{kipf2016semi, hamilton2017inductive, dwivedi2020benchmarking, ying2018graph}. 
Typically it integrate multi-hop neighbors into node representations to capture node feature and graph structure, which hence could model long-range connectivity. KGCN \cite{wang2019knowledge} and KGNN-LS \cite{wang2019knowledge-aware} firstly utilize graph convolutional network (GCN) to obtain item embeddings by aggregating items’ neighborhood information iteratively. Later, KGAT \cite{wang2019kgat} combines user-item graph with knowledge graph as a heterogeneous graph, then utilizes GCN to recursively perform aggregation on it. More recently, KGIN \cite{wang2021learning} models user-item interactions at an intent level, which reveals user intents behind the KG interactions and combines KG interactions to perform GNN on the user-item-entity graph. However, all these approaches adopts the paradigm of supervised learning for model training, relying on their original sparse interactions. 
In contrast, our work explores self-supervised learning in knowledge-aware recommendation, exploiting supervised signals from data itself to improve node representation learning.

\subsection{Contrastive Learning}
Contrastive Learning methods \cite{velickovic2019deep, wu2021self, wang2021self} learn node representations by contrasting positive pairs against negative pairs.
DGI \cite{velickovic2019deep} first adopts Infomax \cite{linsker1988self} in graph representation learning, and focuses on contrasting the local node embeddings with global graph embeddings. Then GMI \cite{peng2020graph} proposes to contrast center node with its local nodes from node features and topological structure. Similarly, MVGRL \cite{hassani2020contrastive} learns node- and graph-level node representations from two structural graph views including first-order neighbors and a graph diffusion, and contrasts encoded embeddings between two graph views. More recently, HeCo \cite{wang2021self} proposes to learn node representations from network schema view and meta-path view, and performs contrastive learning between them. And in traditional collaborative filtering (CF) based recommendation domain, SGL \cite{wu2021self} conducts contrastive learning between original graph and corrupted graph on user-item interactions.
However, little effort has been done towards investigating the great potential of contrastive learning on knowledge-aware recommendation.
\section{Problem Formulation}

In this section, we first introduce two types of necessary structural data, \ie user-item interactions and knowledge graph, and then present the problem statement of our knowledge-aware recommendation problem.

\textbf{Interaction Data}. In a typical recommendation scenario, let $\mathcal{U}=\left\{u_{1}, u_{2}, \ldots, u_{M}\right\}$ and $\mathcal{V}=\left\{v_{1}, v_{2}, \ldots, v_{N}\right\}$ denote the sets of $M$ users and $N$ items, respectively.
The user-item interaction matrix $\mathbf{Y} \in \mathbf{R}^{M \times N}$ is defined according to users’ implicit feedbacks, where $y_{u v}=1$ indicates that user $u$ engaged with item $v$, such as behaviors like clicking or purchasing; otherwise $y_{u v}=0$.

\textbf{Knowledge Graph}.
In addition to the historical interactions, the real-world facts (\eg item attributes, or external commonsense knowledge) associated with items are stored in a KG, in the form of a heterogeneous graph \cite{shi2018heterogeneous,gao2011semi,wei2015ranking}.
Let $\mathcal{G}=\{(h, r, t) \mid h, t \in \mathcal{E}, r \in \mathcal{R}\}$ be the knowledge graph, where $h$, $r$, $t$ are on behalf of head, relation, tail of a knowledge triple correspondingly; $\mathcal{E}$ and $\mathcal{R}$ refer to the sets of entities and relations in $\mathcal{G}$.
For example, the triple (Batman Begins, film.film.star, Christian Bale) means that Christian Bale is an actor of the movie Batman Begins.
In many recommendation scenarios, an item $v \in \mathcal{V}$ corresponds to one entity $e \in \mathcal{E}$. For example, in movie recommendation, the item “Iron Man" also appears in the knowledge graph as an entity with the same name. So we establish a set of item-entity alignments $\mathcal{A} =\{(v, e)|v \in \mathcal{V}, e \in \mathcal{E}\}$, where $\left(v, e\right)$ indicates that item $v$ can be aligned with an entity $e$ in the KG.
With the alignments between items and KG entities, KG is able to profile items and offer complementary information to the interaction data.

\textbf{Problem Statement}. Given the user-item interaction matrix $\mathbf{Y}$ and the knowledge graph $\mathcal{G}$, our task of knowledge-aware recommendation is to learn a function that can predict how likely a user would adopt an item.

\section{Methodology}
\begin{figure*}[th]
  \centering
  \includegraphics[width=\textwidth]{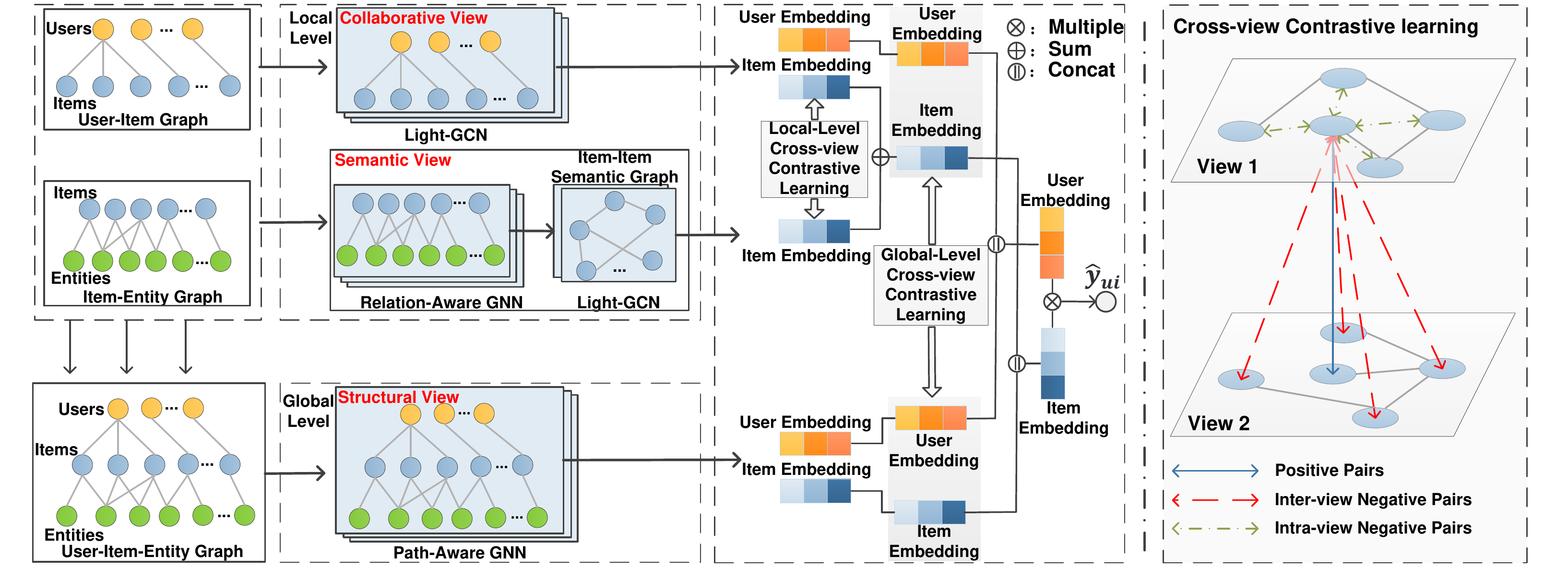}
  \vspace{-0.7cm}
  \caption{Illustration of the proposed MCCLK model. The left subfigure shows model framework of MCCLK; and the right subfigure presents the details of cross-view contrastive learning mechanism. Best viewed in color.}
  \label{fig:model}
  \vspace{-0.2cm}
\end{figure*}

We now present the proposed MCCLK. MCCLK aims to incorporate self-supervised learning into knowledge-aware recommendation for improving user/item representation learning. Figure \ref{fig:model} displays the working flow of MCCLK, which comprises three main components: 
1) \textbf{Multi Views Generation.} It generates three different graph views, including global-level structural view, local-level collaborative and semantic view. For exploring the rarely noticed semantic view, an item-item semantic graph is constructed with a proposed relation-aware GNN.  
2) \textbf{Local-level contrastive learning.} It first encodes collaborative and semantic views with Light-GCN, and then performs cross-view contrastive learning between two views for learning comprehensive node embeddings in the local level.
3) \textbf{Global-level contrastive learning.} It first encodes structural view with path-aware GNN, and then performs cross-view contrastive learning between global- and local-level views for learning discriminative node embeddings in the global level.
We next present the three components in detail.

\subsection{Multi Views Generation}
Different from previous methods only considering global user-item-entity graph, we propose to learn in a more comprehensive and fine-granularity way, by jointly considering local- and global-level view.
We first divide the user-item-entity graph into a user-item graph and an item-entity graph, according to their different types of item-item relationship. 
For the user-item graph, we treat it as \textbf{collaborative view}, aiming to mine the collaborative relationship between items, \ie item-user-item co-occurrences. 
For the item-entity graph, it is viewed as \textbf{semantic view}, towards exploring the semantic similarity between items, \ie item-entity-item co-occurrences.
For the original user-item-entity graph, it is deemed to \textbf{structural view}, aiming to preserve the complete path information, \ie user-item-entity long-range connectivity.

Although much research effort has been devoted to collaborative and structural views, they usually inadequately explore the semantic view, leaving the crucial item-item semantic similarity untouched. 
Towards explicitly considering the item-item semantic relationship, we propose to construct a \textbf{$k$-Nearest-Neighbor item-item semantic graph} $S$ with a relation-aware aggregation mechanism which preserves both neighbor entities and relations information. Each entry $S_{i j}$ in $S$ denotes the semantic similarity between item i and item j. In particular, $S_{i j}=0$ means there is no link between them.

Specifically, we first recursively learn item representations for $K'$ times from the knowledge graph $\mathcal{G}$, with the proposed relation-aware aggregating mechanism as follows:
\begin{equation}
\begin{array}{l}
    \vec{e}_{i}^{(k+1)}=\frac{1}{\left|\mathcal{N}_{i}\right|} \sum\limits_{(r, v) \in \mathcal{N}_{i}} \vec{e}_{r} \odot \vec{e}_{v}^{(k)},\\
     \vec{e}_{v}^{(k+1)}=\frac{1}{\left|\mathcal{N}_{v}\right|}\left( \sum\limits_{(r, v) \in \mathcal{N}_{v}} \vec{e}_{r} \odot \vec{e}_{v}^{(k)} 
     + \sum\limits_{(r, i) \in \mathcal{N}_{v}} \vec{e}_{r} \odot \vec{e}_{i}^{(k)}
     \right),
\end{array}
\end{equation}
where $\vec{e}_{i}^{(k)}$ and $\vec{e}_{v}^{(k)}$  ($\forall k \in K'$) separately denote the representations of item $i$ and entity $v$, which memorize the relational signals propagated from their $(k-1)$-hop neighbors.
For each triplet $(i, r, v)$, a relational message $\vec{e}_{r} \odot \vec{e}_{v}^{(k)}$ is designed for implying different meanings of triplets, via modeling the relation $r$ through the projection or rotation operator \cite{sun2019rotate}.

As such, both neighbor entities and relations in KG are encoded into item representation. Thereafter, inspired by \cite{zhang2021mining}, the item-item similarity graph is built based on a cosine similarity, which is calculated as follows:
\begin{equation}
    S_{i j} =\frac{\left(\vec{e}_{i}^{(K')} \right)^{\top} \vec{e}_{j}^{(K')}}{\left\|\vec{e}_{i}^{(K')}\right\|\left\|\vec{e}_{j}^{(K')}\right\|}.
\end{equation}

Sequentially, a $k$NN sparsification \cite{chen2009fast} is conducted on the fully-connected item-item graph, decreasing computationally demanding, feasible noisy, and unimportant edges \cite{chen2020iterative}.
\begin{equation}
    \widehat{S}_{i j} =\left\{\begin{array}{ll}
S_{i j}, & S_{i j} \in \text { top-k} \left(S_{i}\right), \\
0, & \text {otherwise},
\end{array}\right.
\end{equation}
where $\widehat{S}_{i j}$is a sparsified and directed graph adjacency matrix. To alleviate the exploding or vanishing gradient problem \cite{kipf2016semi}, the adjacency matrix is normalized as follows:
\begin{equation} 
\label{normalize}
    \widetilde{S} =\left(D \right)^{-\frac{1}{2}} \widehat{S} \left(D \right)^{-\frac{1}{2}},
    \vspace{-0.2cm}
\end{equation}
where $D \in \mathbb{R}^{N \times N}$is the diagonal degree matrix of $\widetilde{S}$ and $D_{i, i} = \sum_{j} \widehat{S}_{i j}$. Hence the item-item semantic graph $S$ and its normalized sparsified adjacency matrix $\widetilde{S}$ are finally obtained. 

By doing so, each graph view is now acquired, that is: user-item interaction graph $\mathbf{Y}$ for collaborative view, item-item semantic graph $S$ for semantic view, and the whole user-item-entity graph for structural view. The following local- and global-level contrastive learning are performed across such three views, which will be illustrated in detail.

\subsection{Local-level Contrastive Learning}
Based on the obtained complementary collaborative and semantic views in the local level, we move on to explore two graph views with proper graph encoder, and perform contrastive learning between them to supervise each other. 
Specifically, an effective Light-GCN \cite{he2020lightgcn} is performed in two views to learn a comprehensive item representation.
Then with two view-specific embeddings encoded, the local-level contrastive learning is proposed, encouraging the two views to collaboratively improve representations. 

\subsubsection{Collaborative View Encoder}
The collaborative view stresses collaborative signals between items, \ie item-user-item co-occurrences. As a result, collaborative information could be captured in the collaborative view, by modeling long-range connectivity from user-item interactions. Inspired by precious collaborative filter (CF) based work \cite{he2020lightgcn, wang2019neural}, a Light-GCN is adopted here, which recursively performs aggregation for $K$ times. Light-GCN contains simple message passing and aggregation mechanism without feature transformation and non-linear activation, which is effective and computationally efficient. 
In the $k$-th layer, the aggregation proceeding can be formulated as follows:
\begin{equation}
\begin{array}{l}
\vec{e}_{u}^{(k+1)}=\sum\limits_{i \in \mathcal{N}_{u}} \frac{1}{\sqrt{\left|\mathcal{N}_{u}\right||\mathcal{N} i|}} \vec{e}_{i}^{(k)}, \\
\vec{e}_{i}^{(k+1)}=\sum\limits_{u \in \mathcal{N}_{i}} \frac{1}{\sqrt{\left|\mathcal{N}_{u}\right||\mathcal{N} i|}} \vec{e}_{u}^{(k)},
\end{array}
\end{equation}
where $\vec{e}_{i}^{(k)}$ and $\vec{e}_{u}^{(k)}$ represent embeddings of user $u$ and item $i$ at the $k$-th layer, $\mathcal{N}_{u}, \mathcal{N}_{i}$ represent neighbors of user $u$ and item $i$ respectively. Then we sum representations at different layers up as the local collaborative representations $\vec{z}_i^{c}$ and $\vec{z}_u^{c}$, as follows:
\begin{equation}
    \vec{z}_u^c = \vec{e}_{u}^{(0)}+\cdots+\vec{e}_{u}^{(K)}, \quad \vec{z}_i^c=\vec{e}_{i}^{(0)}+\cdots+\vec{e}_{i}^{(K)}.
\end{equation}

\subsubsection{Semantic View Encoder}
The semantic view focuses on semantic similarity between items, which has been confirmed to be important but ignored by previous work.
Having explicitly constructed the item-item semantic graph from the item-entity affiliations, a Light-GCN is adopted on it with $L$ times aggregation operation, to learn better item representations by injecting item-item affinities into the embedding.
In the $l$-th layer ($\forall l \in L$), the message passing and aggregation process could be formulated as:
\begin{equation}
    \vec{e}_{i}^{(l+1)}=\sum_{j \in \mathcal{N}(i)} \widetilde{S}  \vec{e}_{j}^{(l)},
\end{equation}
where $\mathcal{N}(i)$ is the neighbor items, $\widetilde{S}$ is the normalized sparsified graph adjacency matrix in Equation \ref{normalize}, and $\vec{e}_{i}^{(l)}$ is the $l$-th layer representation of item $i$. Here the input item representation $\vec{e}_{j}^{(0)}$ is set as its corresponding ID embedding vector, rather than the aggregated features, since the Light-GCN is employed in order to directly capture item-item affinities. Then we sum item representations at different layers up to get the local semantic representations $\vec{z}_i^{s}$:
\begin{equation}
    \vec{z}_i^s=\vec{e}_{i}^{(0)}+\cdots+\vec{e}_{i}^{(L)}.
\end{equation}

\subsubsection{Local-level Cross-view Contrastive Optimization}
With the view-specific embeddings $\vec{z}_i^{s}$ and $\vec{z}_i^{c}$ for item $i$ from the collaborative and semantic views, a local-level cross-view contrastive learning is performed, for supervising two views to learn discriminative representations. 
Aiming to map them into the space where contrastive loss is calculated, embeddings are first feed into a MLP with one hidden layer:
\begin{equation}
    \begin{array}{c}
{\vec{z}_{i}^{c}}_{-} \text {p}=W^{(2)} \sigma\left(W^{(1)} \vec{z}_{i}^{c}+b^{(1)}\right)+b^{(2)}, \\
{\vec{z}_{i}^{s}}_{-} \text {p}=W^{(2)} \sigma\left(W^{(1)} \vec{z}_{i}^{s}+b^{(1)}\right)+b^{(2)},
\end{array}
\end{equation}
where $W^{(\cdot)} \in \mathbb{R}^{d \times d}$ and $b^{(\cdot)} \in \mathbb{R}^{d \times 1}$ are trainable parameters, $\sigma$ is ELU non-linear function. 
Then we define the positive and negative samples here, inspired by works in other areas \cite{zhu2020deep, zhu2021graph}, for any node in one view, the same node embedding learned by the other view forms the positive sample; and in two views, nodes embeddings other than it are naturally regarded as negative samples.

With the defined positive and negative samples, we have the following contrastive loss:
\begin{equation}
\begin{array}{ll}
    \mathcal{L}^{local}= 
    -\log \frac{e^{\operatorname{s}\left({{\vec{z}_{i}^{s}}_{-} \text {p}}, { {\vec{z}_{i}^{c}}_{-} \text {p}}\right) / \tau }}{ \underbrace{e^{ \operatorname{s}\left({ {\vec{z}_{i}^{s}}_{-} \text {p}}, {{\vec{z}_{i}^{c}}_{-} \text {p}}\right) / \tau }}_{\text {positive pair }} +
    \underbrace{\sum\limits_{k \neq i}e^{\operatorname{s}\left({{\vec{z}_{i}^{s}}_{-} \text {p}}, {{\vec{z}_{k}^{s}}_{-} \text {p}} \right)/ \tau } }_{\text {intra-view negative pairs }}+
    \underbrace{\sum\limits_{k \neq i} e^{\operatorname{s}\left( {{\vec{z}_{i}^{s}}_{-} \text {p}}, { {\vec{z}_{k}^{c}}_{-} \text {p}}\right) / \tau}}_{\text {inter-view negative pairs}}},
    \end{array}
\label{loss}
\end{equation}
where $\operatorname{s}(\cdot)$ denotes the cosine similarity calculating, and $\tau$ denotes a temperature parameter. 
It's worth mentioning that negative samples come from two sources, which are intra-view and inter-view nodes, corresponding to the second and the third term in the denominator in Equation \ref{loss}.
In this way, the local-level cross-view contrastive learning is successfully achieved.

\subsection{Global-level Contrastive Learning}
Although user/item feature information has been revealed from local-level views, the complete graph structural information hasn't been explored, that is, the long-range connectivity unifying both user-item and item-entity graphs, \ie user-item-entity connections.
Hence the global-level contrastive learning is introduced, which first explores the structural view with a path-aware encoder, and then performs contrastive learning between the global-level and local-level views to supervise each other level. 
To be more specific, inspired by \cite{wang2021learning}, we design a path-aware GNN to automatically encode path information into node embeddings.
Then with the encoded embeddings from global-level view and local-level view, the global-level contrastive learning is performed, for supervising two-level views to learn comprehensive representations.

\subsubsection{Structural View Encoder}
Aiming to encode the structural information under structural view (\ie the variety of paths), inspired by \cite{wang2021learning}, a path-aware GNN is proposed here, which aggregates neighboring information for $L'$ times meanwhile preserving the path information, \ie long-range connectivity such as user-interact-item-relation-entity.

In particular, in $l$-th layer $(\forall l \in L')$ the aggregation process can be formulated as:
\begin{equation}
    \begin{array}{l}
\vec{e}_{u}^{(l+1)}=\frac{1}{\left|\mathcal{N}_{u}\right|} \sum\limits_{i \in \mathcal{N}_{u}} \vec{e}_{i}^{(l)}, \\
\vec{e}_{i}^{(l+1)}=\frac{1}{\left|\mathcal{N}_{i}\right|} \sum\limits_{(r, v) \in \mathcal{N}_{i}} \beta(i, r, v) \vec{e}_{r} \odot \vec{e}_{v}^{(l)},
\end{array}
\end{equation}
where $\vec{e}_{i}^{(l)}$ and $\vec{e}_{v}^{(l)}$ separately denote the representations of item $i$ and entity $v$, which memorize the relational signals propagated from their $(l-1)$-hop neighbors and hence store the holistic semantics of multi-hop paths. And aiming to weight each relation and entity, the attention weights $\beta(i, r, v)$ is calculated as follows:
\begin{equation}
    \begin{aligned}
\beta(i, r, v) &=\operatorname{softmax}\left(\left(\vec{e}_{i} || \vec{e}_{r}\right)^{T} \cdot \left(\vec{e}_{v} || \vec{e}_{r}\right) \right) \\
&=\frac{\exp \left(\left(\vec{e}_{i} || \vec{e}_{r}\right)^{T} \cdot \left(\vec{e}_{v} || \vec{e}_{r}\right)\right)}{\sum\limits_{\left(v^{\prime}, r\right) \in \hat{N}(i)} \exp \left(\left(\vec{e}_{i}|| \vec{e}_{r}\right)^{T} \cdot \left(\vec{e}_{v^{\prime}} || \vec{e}_{r}\right)\right)},
    \end{aligned}
\end{equation}
where $||$ denotes concat operation, $\hat{N}(i)$ denotes the set of neighboring entities $N(i)$ and item $i$ itself. Then we sum all layers' representations up to have the global representations $\mathbf{z}_u^{g}$ and $\mathbf{z}_i^{g}$ :
\begin{equation}
    \vec{z}_u^g = \vec{e}_{u}^{(0)}+\cdots+\vec{e}_{u}^{(L')}, \quad \vec{z}_i^g=\vec{e}_{i}^{(0)}+\cdots+\vec{e}_{i}^{(L')}.
\end{equation}

\subsubsection{Global-level Cross-view Contrastive Optimization}
Obtaining the node representations under global- and local-level views, they are first mapped into the space where the contrastive loss is calculated, the same as local-level contrastive loss calculating:
\begin{equation}
    \begin{array}{c}
{\vec{z}_{i}^{g}}_{-} \text {p}=W^{(2)} \sigma\left(W^{(1)} \vec{z}_{i}^{g}+b^{(1)}\right)+b^{(2)}, \\
{\vec{z}_{i}^{l}}_{-} \text {p}=W^{(2)} \sigma\left(W^{(1)} (\vec{z}_{i}^{c}+\vec{z}_{i}^{s})+b^{(1)}\right)+b^{(2)}.
\end{array}
\end{equation}

With the same positive and negative sampling strategy to local-level contrastive learning, we have the following contrastive loss:
\begin{equation}
\begin{array}{l}
    \mathcal{L}_{i}^{g}= 
    -\log \frac{e^{\operatorname{s}\left({{\vec{z}_{i}^{g}}_{-} \text {p}}, { {\vec{z}_{i}^{l}}_{-} \text {p}}\right) / \tau }}{ \underbrace{e^{ \operatorname{s}\left({ {\vec{z}_{i}^{g}}_{-} \text {p}}, {{\vec{z}_{i}^{l}}_{-} \text {p}}\right) / \tau }}_{\text {positive pair }} +
    \underbrace{\sum\limits_{k \neq i}e^{\operatorname{s}\left({{\vec{z}_{i}^{g}}_{-} \text {p}}, {{\vec{z}_{k}^{g}}_{-} \text {p}} \right)/ \tau } }_{\text {intra-view negative pairs }}+
    \underbrace{\sum\limits_{k \neq i} e^{\operatorname{s}\left( {{\vec{z}_{i}^{g}}_{-} \text {p}}, { {\vec{z}_{k}^{l}}_{-} \text {p}}\right) / \tau}}_{\text {inter-view negative pairs}}},  \\
    \mathcal{L}_{i}^{l}= 
    -\log \frac{e^{\operatorname{s}\left({{\vec{z}_{i}^{l}}_{-} \text {p}}, { {\vec{z}_{i}^{g}}_{-} \text {p}}\right) / \tau }}{ \underbrace{e^{ \operatorname{s}\left({ {\vec{z}_{i}^{l}}_{-} \text {p}}, {{\vec{z}_{i}^{g}}_{-} \text {p}}\right) / \tau }}_{\text {positive pair }} +
    \underbrace{\sum\limits_{k \neq i}e^{\operatorname{s}\left({{\vec{z}_{i}^{l}}_{-} \text {p}}, {{\vec{z}_{k}^{l}}_{-} \text {p}} \right)/ \tau } }_{\text {intra-view negative pairs }}+
    \underbrace{\sum\limits_{k \neq i} e^{\operatorname{s}\left( {{\vec{z}_{i}^{l}}_{-} \text {p}}, { {\vec{z}_{k}^{g}}_{-} \text {p}}\right) / \tau}}_{\text {inter-view negative pairs}}},  \\
    \end{array}
\end{equation}
where $\mathcal{L}_{i}^{g}$ and $\mathcal{L}_{i}^{l}$ denote the contrastive learning loss calculated from the global view and local view. And the contrastive loss $\mathcal{L}_{u}^{g}$/$\mathcal{L}_{u}^{l}$ calculating from user embedding is similar as $\mathcal{L}_{i}^{g}$/$\mathcal{L}_{i}^{l}$, where only item embeddings are exchanged into user embeddings in the formula. Then the overall objective is given as follows:
\begin{equation}
\mathcal{L}^{global} = \frac{1}{2 N} \sum\limits_{i=1}^{N}(\mathcal{L}_{i}^{g} + \mathcal{L}_{i}^{l}) + 
\frac{1}{2 M} \sum\limits_{i=1}^{M}(\mathcal{L}_{u}^{g} + \mathcal{L}_{u}^{l}).
\end{equation}

\subsection{Model Prediction}
After performing multi-layer aggregation in three views and optimizing through multi-level cross-view contrastive learning, we obtain multiple representations for user $u$, namely $z_u^c$ and $z_u^g$; analogous to item $i$, $z_i^c$, $z_i^s$ and $z_i^g$. By summing and concatenating the above representations, we have the final user/item representations and predict their matching score through inner product, as follows:
\begin{equation}
\begin{array}{l}
    \vec{z}_{u}^{*}=\vec{z}_{u}^{g} || \vec{z}_{u}^{c},\\
    \vec{z}_{i}^{*}=\vec{z}_{i}^{g} || (\vec{z}_{i}^{c}+\vec{z}_{i}^{s}),\\
    \hat{y}(u, i)=\vec{z}_{u}^{* \top} \vec{z}_{i}^{*}.
\end{array}
\end{equation}

\subsection{Multi-task Training}
To combine the recommendation task with the self-supervised task, we optimize the whole model with a multi-task training strategy. For the KG-aware recommendation task, a pairwise BPR loss \cite{rendle2012bpr} is adopted to reconstruct the historical data, which encourages the prediction scores of a user’s historical items to be higher than the unobserved items.
\begin{equation}
    \mathcal{L}_{\mathrm{BPR}}=\sum_{(u, i, j) \in O}-\ln \sigma\left(\hat{y}_{u i}-\hat{y}_{u j}\right),
\end{equation}
where $\boldsymbol{O}=\left\{(u, i, j) \mid(u, i) \in \boldsymbol{O}^{+},(u, j) \in \boldsymbol{O}^{-}\right\}$ is the training dataset consisting of the observed interactions $\boldsymbol{O}^{+}$ and unobserved counterparts $\boldsymbol{O}^{-}$; $\sigma$ is the sigmoid function. By combining the global- and local-level contrastive loss with BPR loss, we minimize the following objective function to learn the model parameter:
\begin{equation}
     \mathcal{L}_{MCCLK} = \mathcal{L}_{\mathrm{BPR}} +  \beta (\alpha\mathcal{L}^{local} + (1-\alpha)\mathcal{L}^{global}) + \lambda\|\Theta\|_{2}^{2},
\end{equation}
where $\Theta$ is the model parameter set, $\alpha$ is a hyper parameter to determine the local-global contrastive loss ratio, $\beta$ and $\lambda$ are two hyper parameters to control the contrastive loss and $L_2$ regularization term, respectively.
\section{Experiment}

\begin{table}[t]
\centering
\setlength{\tabcolsep}{0.8mm}{
\begin{tabular}{cl|ccc}
\hline
\multicolumn{1}{l}{}                                                                &                 & \multicolumn{1}{l}{Book-Crossing} & \multicolumn{1}{l}{MovieLens-1M} & \multicolumn{1}{l}{Last.FM} \\ \hline \hline
\multirow{3}{*}{\begin{tabular}[c]{@{}c@{}}User-item   \\ Interaction\end{tabular}} & \# users  & 17,860 & 6,036 & 1,872                       \\
& \# items & 14,967 & 2,445 & 3,846                       \\
& \# interactions & 139,746 & 753,772 & 42,346                      \\ \hline
\multirow{3}{*}{\begin{tabular}[c]{@{}c@{}}Knowledge\\ Graph\end{tabular}} & \# entities & 77,903 & 182,011 & 9,366                       \\
& \# relations & 25 & 12 & 60                          \\
& \# triplets & 151,500 & 1,241,996 & 15,518                      \\ \hline
\multirow{6}{*}{\begin{tabular}[c]{@{}c@{}}Hyper-\\parameter\\ Settings\end{tabular}} 
& \# $\alpha$  & 0.2 & 0.2  & 0.2                      \\
& \# $\beta$  & 0.1 & 0.1  & 0.1                      \\
& \# $K$ & 2 & 2 & 3                       \\
& \# $K'$ & 2 & 2 & 2                          \\
& \# $L$ & 1 & 1 & 2                           \\
& \# $L'$& 2 & 2 & 2                          \\ \hline
\end{tabular}}
\caption{ Statistics and hyper-parameter settings for the three datasets.($\alpha$: local-level contrastive loss weight, $\beta$: contrastive loss weight, $K$: local collaborative aggregation depth, $K'$: aggregation depth of item-item semantic graph construction, $L$: local semantic aggregation depth, $L'$: global structural aggregation depth.)}
\label{datasets}
\end{table}

\begin{table*}[t]
    \centering
    \setlength{\tabcolsep}{8pt}
    \begin{tabular}{c|ll|ll|ll}
    	\hline
    	\multirow{2}{*}{Model} & \multicolumn{2}{c}{Book-Crossing} & \multicolumn{2}{c}{MovieLens-1M} & \multicolumn{2}{c}{Last.FM} \\
        & \multicolumn{1}{c}{\textit{AUC}} & \multicolumn{1}{c}{\textit{F1}} & \multicolumn{1}{c}{\textit{AUC}} & \multicolumn{1}{c}{\textit{F1}} & \multicolumn{1}{c}{\textit{AUC}} & \multicolumn{1}{c}{\textit{F1}} \\
        \hline\hline
BPRMF & 0.6583$(-10.42\%)$ & 0.6117$(-6.60\%)$ & 0.8920$(-4.31\%)$ & 0.7921$(-7.10\%)$ & 0.7563$(-12.00\%)$ & 0.7010$(-9.98\%)$    \\
\hline
CKE   & 0.6759$(-8.66\%)$ & 0.6235$(-5.42\%)$ & 0.9065$(-2.86\%)$ & 0.8024$(-6.07\%)$ & 0.7471$(-12.92\%)$ & 0.6740$(-12.68\%)$  \\
RippleNet   & 0.7211$( -4.14\% )$ & 0.6472$( -3.05\% )$ & 0.9190$(-1.61\% )$ & 0.8422$( -2.09\% )$ & 0.7762$( -10.01\% )$ & 0.7025$( -9.83\% )$ \\
\hline
PER   & 0.6048$(-15.77\%)$ & 0.5726$(-10.51\%)$ & 0.7124$(-22.27\%)$ & 0.6670$(-19.61\%)$ & 0.6414$(-23.49\%)$ & 0.6033$(-19.75\%)$          \\
\hline
KGCN     & 0.6841$(-7.84\%)$ & 0.6313$(-4.64\%)$  & 0.9090$(-2.61\%)$ & 0.8366$(-2.65\%)$ & 0.8027$(-7.36\%)$ & 0.7086$(-9.22\%)$          \\
KGNN-LS    & 0.6762$(-8.63\%)$ & 0.6314$(-4.63\%)$ & 0.9140$(-2.11\%)$ & 0.8410$(-2.21\%)$ & 0.8052$(-7.11\%)$ & 0.7224$(-7.84\%)$          \\
KGAT    & \underline{0.7314}$(-3.11\%)$ & 0.6544$(-2.33\%)$ & 0.9140$(-2.11\%)$ & 0.8440$(-1.91\%)$ & 0.8293$(-4.70\%)$ & 0.7424$(-5.84\%)$          \\
KGIN    & 0.7273$(-3.52\%)$   & \underline{0.6614}$(-1.63\%)$ & \underline{0.9190}$(-1.61\%)$ & \underline{0.8441}$(-1.90\%)$ & \underline{0.8486}$(-2.77\%)$ & \underline{0.7602}$(-4.06\%)$ \\ \hline
\textbf{MCCLK}    & \textbf{0.7625}* & \textbf{0.6777}* & \textbf{0.9351}* & \textbf{0.8631}* & \textbf{0.8763}* & \textbf{0.8008}*           \\ \hline
\end{tabular}
\caption{The result of $AUC$ and $F1$ in CTR prediction. The best results are in boldface and the second best results are underlined. * denotes statistically significant improvement by unpaired two-sample $t$-test with $p < 0.001$.}
\label{tab:compare}
\end{table*}

\begin{figure*}[htbp]  
    \centering  
    \subfloat[Book-Crossing] 
    {   \begin{minipage}[t]{0.33\textwidth}
            \centering     
            \includegraphics[width=\textwidth]{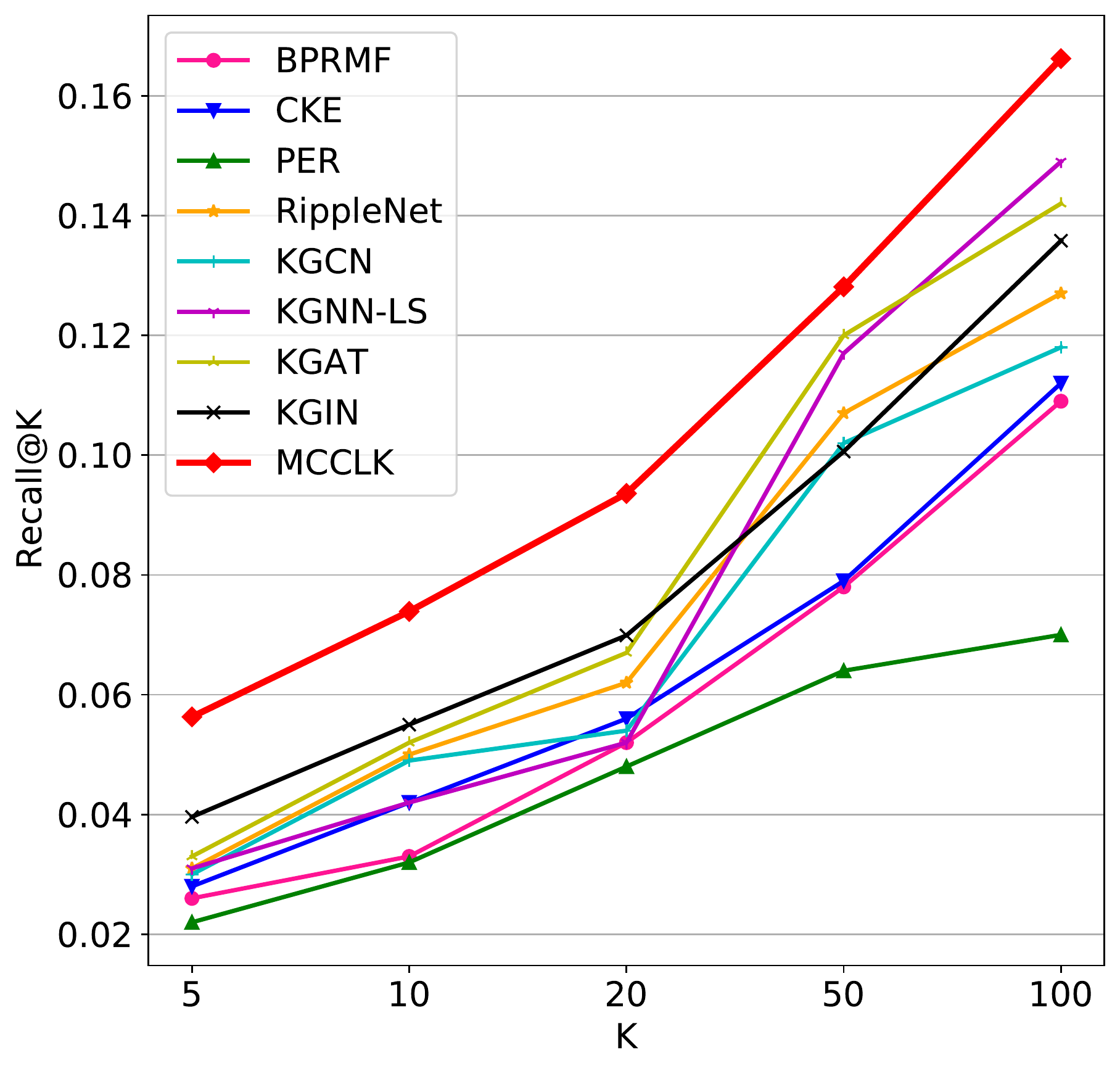} 
        \end{minipage}}
    \subfloat[MovieLens-1M] 
    {
        \begin{minipage}[t]{0.33\textwidth}
            \centering     
            \includegraphics[width=\textwidth]{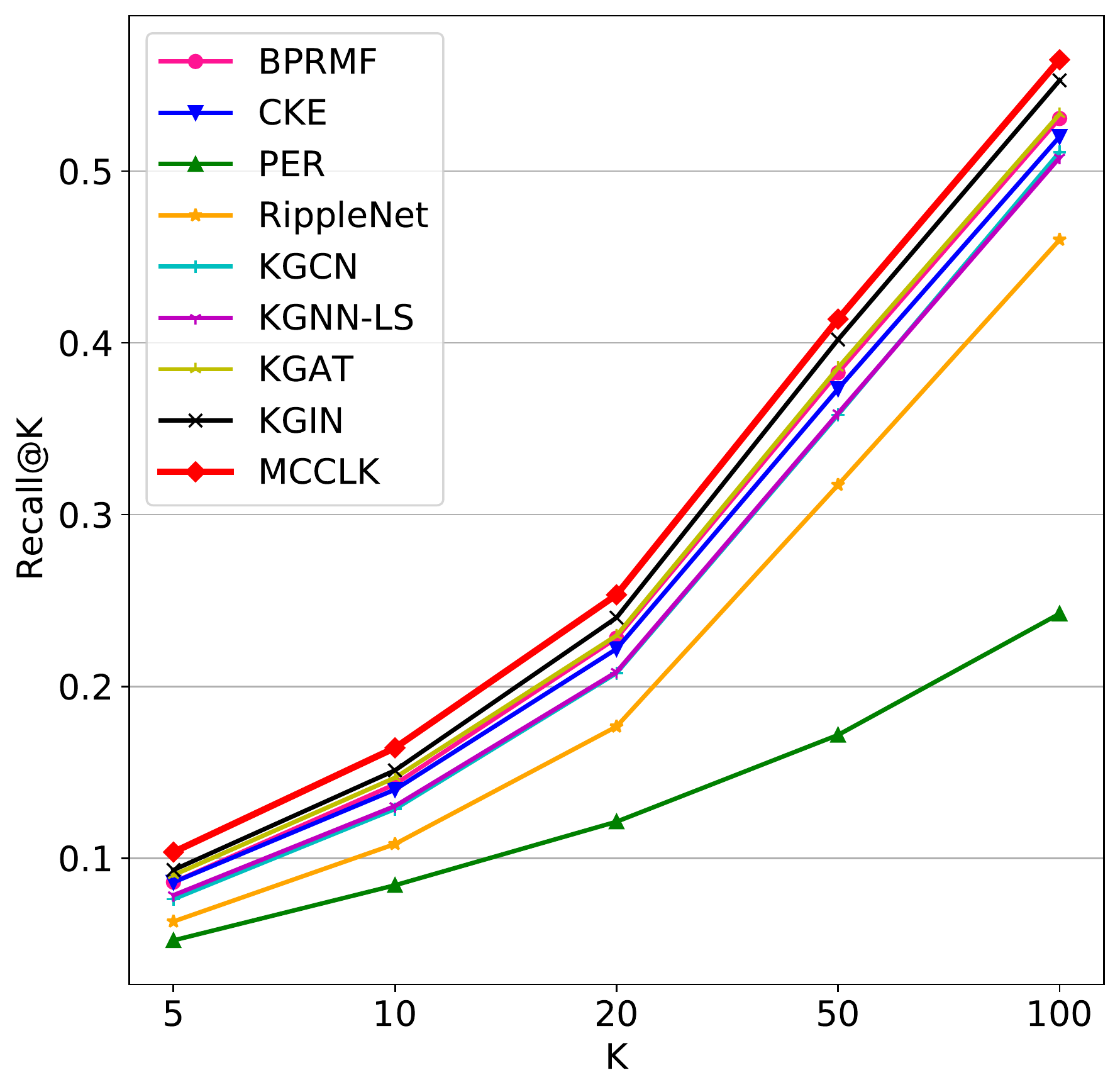}   
        \end{minipage}
    }%
    \subfloat[Last.FM] 
    {
        \begin{minipage}[t]{0.33\textwidth}
            \centering      
            \includegraphics[width=\textwidth]{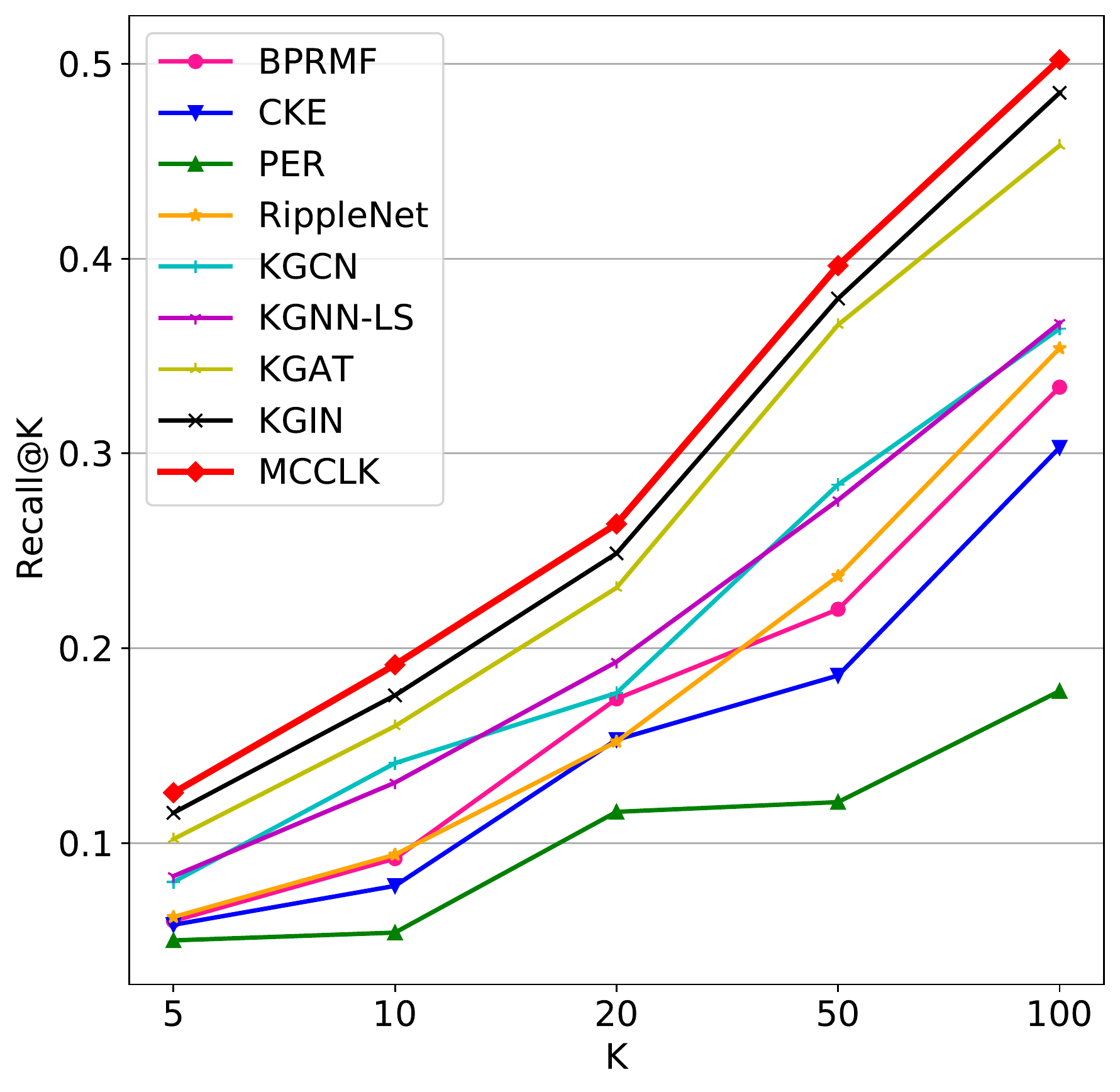}   
        \end{minipage}
    }%
    \vspace{-0.3cm}
    \caption{The result of Recall@$K$ in top-$K$ recommendation.} 
    \label{fig::topk} 
\end{figure*}

Aiming to answer the following research questions, we conduct extensive
experiments on three public datasets:
\begin{itemize}
    \item \textbf{RQ1:} How does MCCLK perform, compared to present models?
    \item \textbf{RQ2:} Are the main components (\eg local-level contrastive learning, global-level contrastive learning) really working well?
    \item \textbf{RQ3:} How do different hyper-parameter settings (\eg aggregation layer in structural view, local-level contrastive loss weight $\alpha$ \etc) affect MCCLK?
    \item \textbf{RQ4:} Is the self-supervised task really improving the representation learning?
\end{itemize}

\subsection{Experiment Settings}
\subsubsection{Dataset Description}

We use three benchmark datasets to evaluate the effectiveness of MCCLK: Book-Crossing, MovieLens-1M, and Last.FM. The three datasets of different domains are publicly accessible and vary in size and sparsity, making our experiments more convincing.
\begin{itemize}
\item{\verb|Book-Crossing|\footnote{\url{http://www2.informatik.uni-freiburg.de/~cziegler/BX/}}}: It is collected from the book-crossing community, which consists of trenchant ratings (ranging from 0 to 10) about various books.
\item {\verb|MovieLens-1M|\footnote{\url{https://grouplens.org/datasets/movielens/1m/}}}: It’s a benchmark dataset for movie recommendations, which contains approximately 1 million explicit ratings (ranging from 1 to 5) on a total of 2,445 items from 6,036 users.
\item{\verb|Last.FM|\footnote{\url{https://grouplens.org/datasets/hetrec-2011/}}}: It is a music listening dataset collected from Last.FM online music systems with around 2 thousand users.
\end{itemize}

Since the interactions in MovieLens-1M, Book-Crossing, and Last.FM are explicit feedback, we follow RippleNet \cite{wang2018ripplenet} and transform them into the implicit feedback in which 1 indicates the positive samples (the threshold of the rating to be viewed as positive is 4 for MovieLens-1M, but no threshold is set for Last.FM and Book-Crossing due to their sparsity). As to negative samples, for each user, we randomly sample from his unwatched items with the size equal to his positive ones.

As for the sub-KG construction, we follow RippleNet \cite{wang2018ripplenet} and use Microsoft Satori\footnote{\url{https://searchengineland.com/library/bing/bing-satori}} to construct it for MovieLens-1M, Book-Crossing, and Last.FM datasets. Each sub knowledge graph that follows the triple format is a subset of the whole KG with a confidence level greater than 0.9. Given the sub-KG, we gather Satori IDs of all valid movies/books/musicians through matching their names with the tail of triples. Then we match the item IDs with the head of all triples and select all well-matched triples from the sub-KG. The basic statistics of the three datasets are presented in \autoref{datasets}.

\subsubsection{Baselines}
To demonstrate the effectiveness of our proposed MCCLK, we compare MCCLK with four types of recommender system methods: CF-based methods (BPRMF), embedding-based method (CKE, RippleNet), path-based method (PER), and GNN-based methods(KGCN, KGNN-LS, KGAT, KGIN) as follows:
\begin{itemize}
\item {\verb|BPRMF| \cite{rendle2012bpr}}: It’s a typical CF-based method that uses pairwise matrix factorization for implicit feedback optimized by the BPR loss.
\item{\verb|CKE| \cite{zhang2016collaborative}}: It’s a embedding-based method that combines structural, textual, and visual knowledge in one framework.
\item {\verb|RippleNet| \cite{wang2018ripplenet}}: It’s a classical embedding-based method which propagates users’ preferences on the KG.
\item{\verb|PER| \cite{yu2014personalized}}: It’s a typical path-based method which extracts meta-path-based features to represent the connectivity between users and items.
\item {\verb|KGCN| \cite{wang2019knowledge}}: It’s a GNN-based method which iteratively integrates neighboring information to enrich item embeddings.
\item {\verb|KGNN-LS| \cite{wang2019knowledge-aware}}: It is a GNN-based model which enriches item embeddings with GNN and label smoothness regularization.
\item {\verb|KGAT| \cite{wang2019kgat}}: It’s a GNN-based method which iteratively integrates neighbors on user-item-entity graph with an attention mechanism to get user/item representations.
\item {\verb|KGIN| \cite{wang2021learning}}: It's a state-of-the-art GNN-based method, which disentangles user-item interactions at the granularity of user intents, and performs GNN on the proposed user-intent-item-entity graph.
\end{itemize}

\subsubsection{Evaluation Metrics}
We evaluate our method in two experimental scenarios: (1) In click-through rate (CTR) prediction, we apply the trained model to predict each interaction in the test set. We adopt two widely used metrics \cite{wang2018ripplenet, wang2019knowledge} $AUC$ and $F1$ to evaluate CTR prediction. (2) In top-$K$ recommendation, we use the trained model to select $K$ items with the highest predicted click probability for each user in the test set, and we choose Recall@$K$ to evaluate the recommended sets.

\subsubsection{Parameter Settings}
We implement our MCCLK and all baselines in Pytorch and carefully tune the key parameters. For a fair comparison, we fix the embedding size to 64 for all models, and the embedding parameters are initialized with the Xavier method \cite{glorot2010understanding}. We optimize our method with Adam \cite{kingma2014adam} and set the batch size to 2048. A grid search is conducted to confirm the optimal settings, we tune the learning rate $\eta$ among$\{0.0001, 0.0003,0.001,0.003\}$ and $\lambda$ of $L2$ regularization term among $\{10^{-7}, 10^{-6}, 10^{-5}, 10^{-4}, 10^{-3}\}$. Other hyper-parameter settings are provided in Table~\ref{datasets}. The best settings for hyper-parameters in all comparison methods are researched by either empirical study or following the original papers.

\subsection{Performance Comparison (RQ1)}

We report the empirical results of all methods in Table \ref{tab:compare} and Figure \ref{fig::topk}. The improvements and statistical significance test are performed between MCCLK and the strongest baselines (highlighted with underline). Analyzing such performance comparison, we have the following observations:
\begin{itemize}
    \item \textbf{Our proposed MCCLK achieves the best results.}
    MCCLK consistently outperforms all baselines across three datasets in terms of all measures. More specifically, it achieves significant improvements over the strongest baselines \wrt AUC by 3.11\%, 1.61\%, and 2.77\% in Book, Movie, and Music respectively, which demonstrates the effectiveness of MCCLK.
    We attribute such improvements to the following aspects: 
    (1) By contrasting collaborative and semantic views at the local level, MCCLK is able to capture collaborative and semantic feature information better;
    (2) The global-level contrastive mechanism preserves both structure and feature information from two-level self-discrimination, hence capturing more comprehensive information for MCCLK than methods only modeling global structure.
    \item \textbf{Incorporating KG benefits recommender system.}
    Comparing CKE with BPRMF, leaving KG untapped limits the performance of MF. By simply incorporating KG embeddings into MF, CKE performs better than MF. Such findings are consistent with prior studies \cite{cao2019unifying}, indicating the importance of side information like KG.
    \item \textbf{The way of exploiting KG information determines the model performance.}
    Path-based method PER performs even worse than BPRMF, because the optimal user-defined meta-paths are hard to be defined in reality. This fact stresses the importance of capturing structural path information from the whole graph. 
    \item \textbf{GNN has a strong power of graph learning.}
    Most of the GNN-based methods perform better than embedding bassed and path-based ones, suggesting the importance of modeling long-range connectivity for graph representation learning. The truth inspires us that learning local/global graph information with a proper aggregation mechanism could improve the model performance.
    
\end{itemize}

\subsection{Ablation Studies (RQ2)}

\begin{figure}[t] 
    \centering  
    \subfloat[Book]
    {
        \begin{minipage}[t]{0.33\linewidth}
            \centering          
            \includegraphics[width=\textwidth]{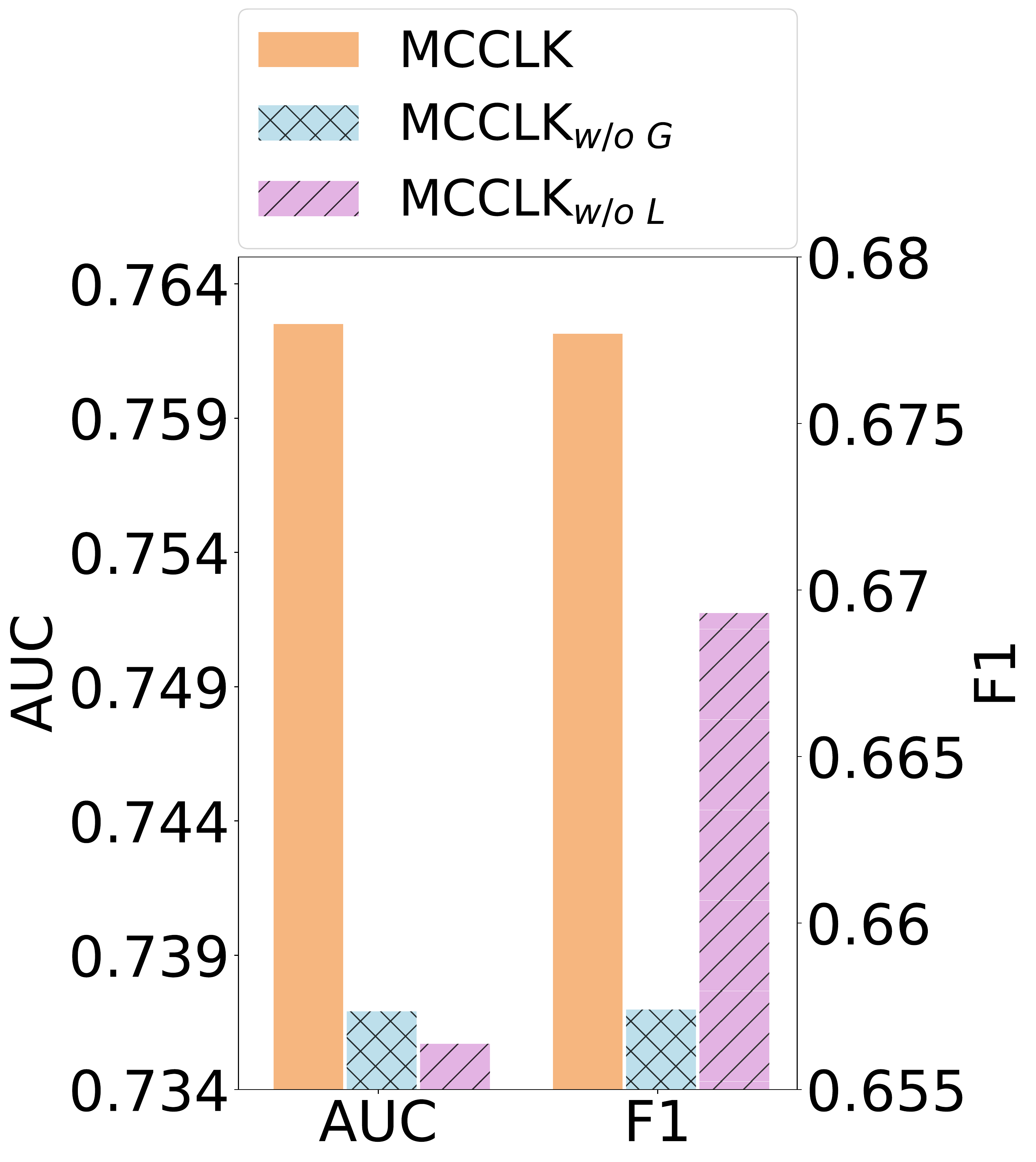}
        \end{minipage}
    }
    \subfloat[Movie]
    {
        \begin{minipage}[t]{0.33\linewidth}
            \centering     
            \includegraphics[width=\textwidth]{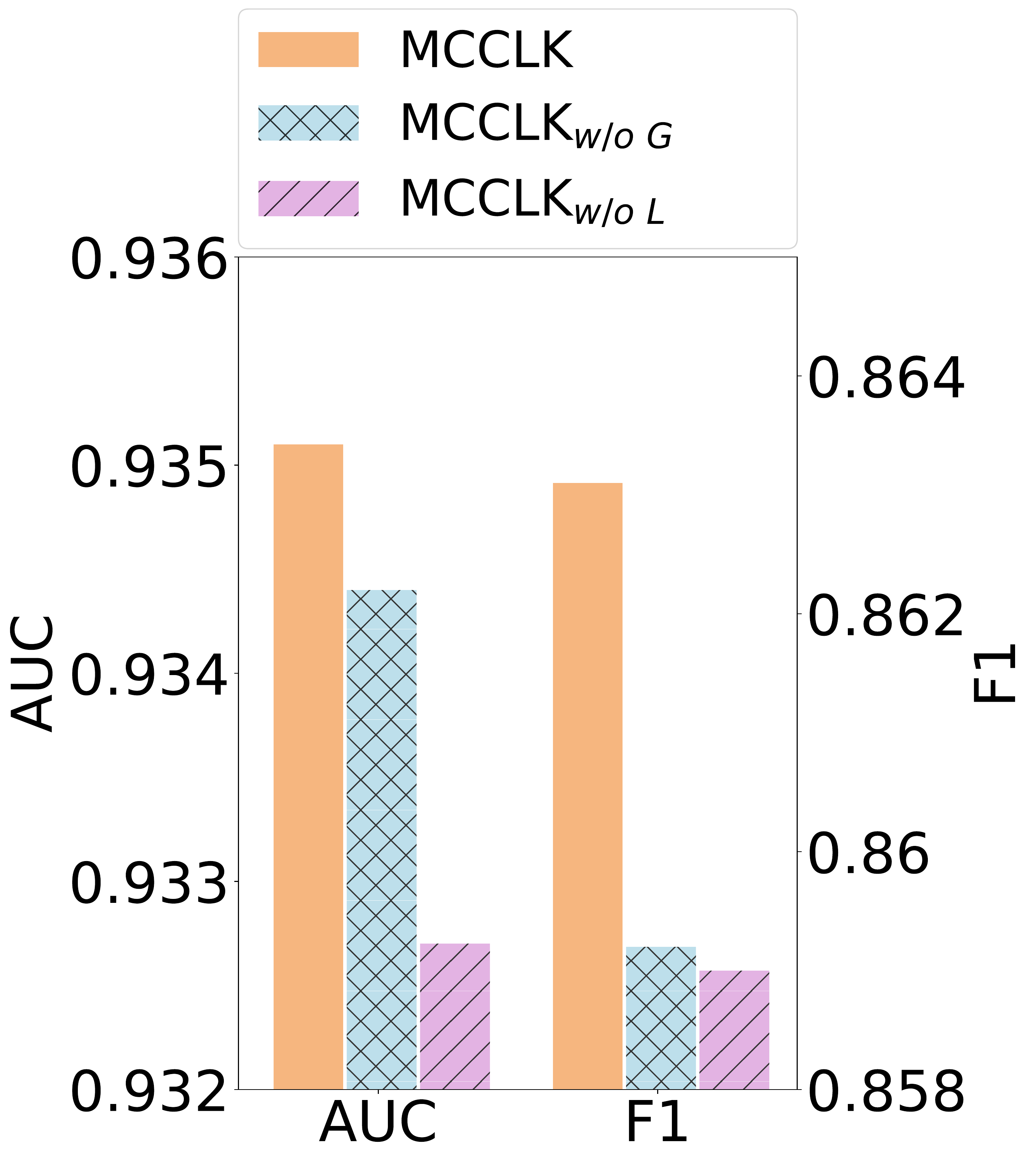}   
        \end{minipage}
    }%
    \subfloat[Music]
    {
        \begin{minipage}[t]{0.33\linewidth}
            \centering      
            \includegraphics[width=\textwidth]{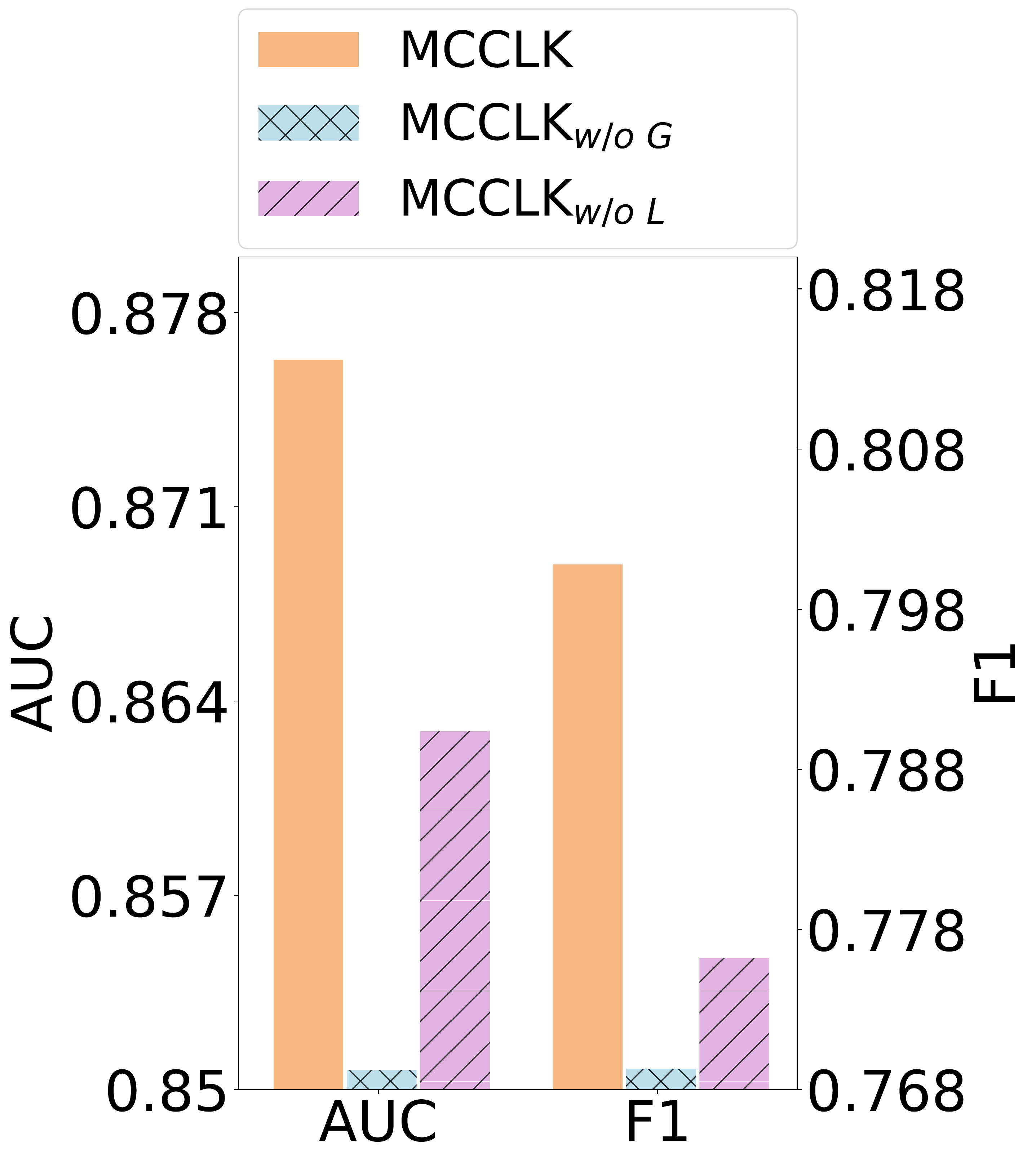}   
        \end{minipage}
    }%
    \caption{Effect of ablation study.} 
    \label{fig::ablation} 
\end{figure}


As shown in table \ref{fig::ablation}, here we examine the contributions of main components in our model to the final performance by comparing MCCLK with the following two variants:
\begin{itemize}
    \item $\text{MCCLK}_{w/o \ G}$ : In this variant, the global-level contrastive learning module is removed, nodes are encoded from two local level views.
    \item $\text{MCCLK}_{w/o \ L}$: This variant removes the local-level contrastive learning module, and only remains the structural view learning of the user-bundle-item graph.
\end{itemize}
The results of two variants and MCCLK are reported in Figure \ref{fig::ablation}, from which we have the following observations: 
1) Removing the global-level contrastive learning significantly degrades the model's performance, which suggests its importance of exploring graph structural information for KG-aware recommendation.
2) In most cases, $\text{MCCLK}_{w/o \ L}$ is the least competitive model, especially in massive datasets (\ie book and movie), which demonstrates the superiority of learning discriminative information between collaborative and semantic views in the local level.

\subsection{Sensitivity Analysis (RQ3)}




\begin{figure*}[t] 
    \centering  
    \subfloat[MCCLK]
    {   \begin{minipage}[t]{0.19\linewidth}
            \centering       
            \includegraphics[width=\textwidth]{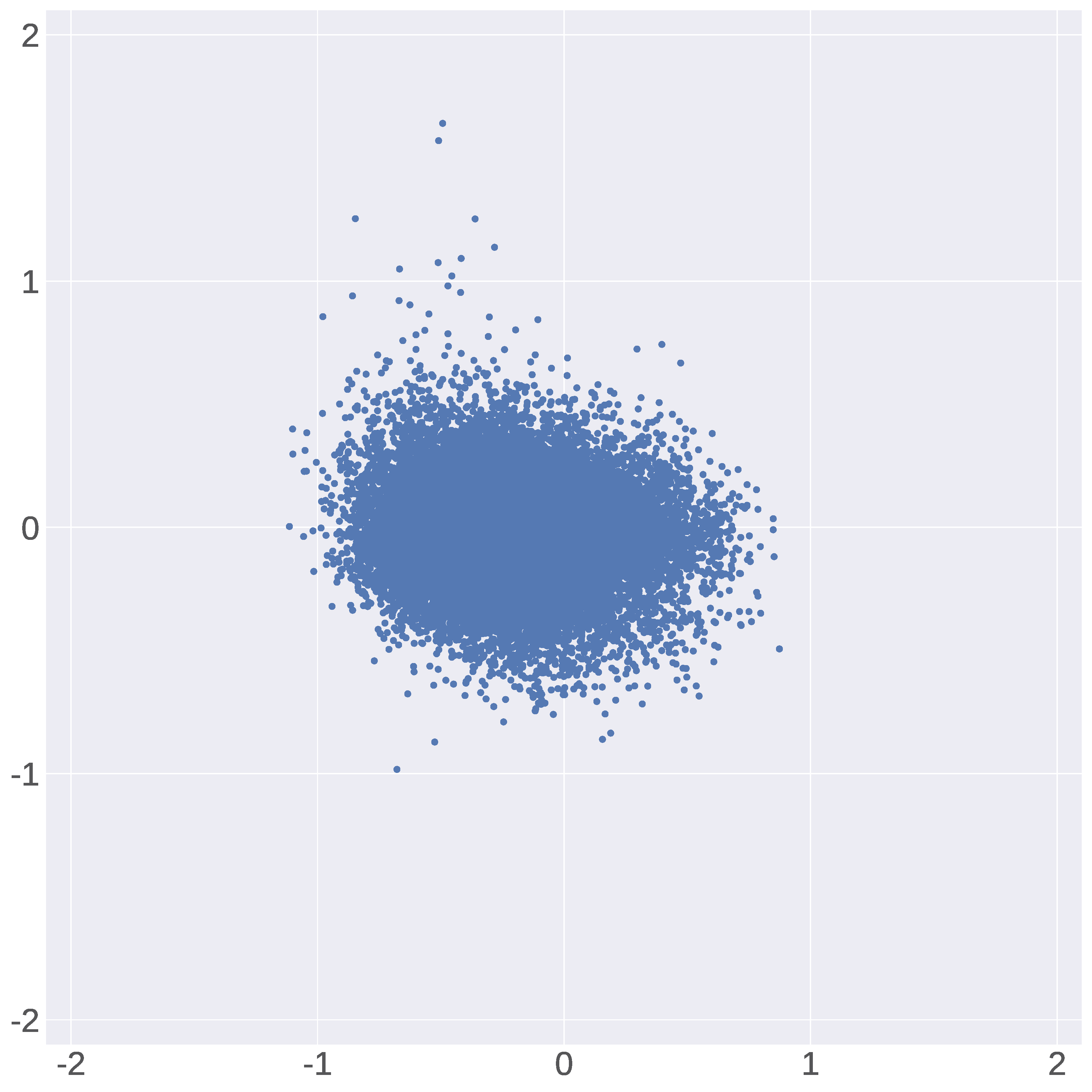} 
        \end{minipage}}
    \subfloat[$\text{MCCLK}_{w/o \ G}$]
    {   \begin{minipage}[t]{0.19\linewidth}
            \centering     
            \includegraphics[width=\textwidth]{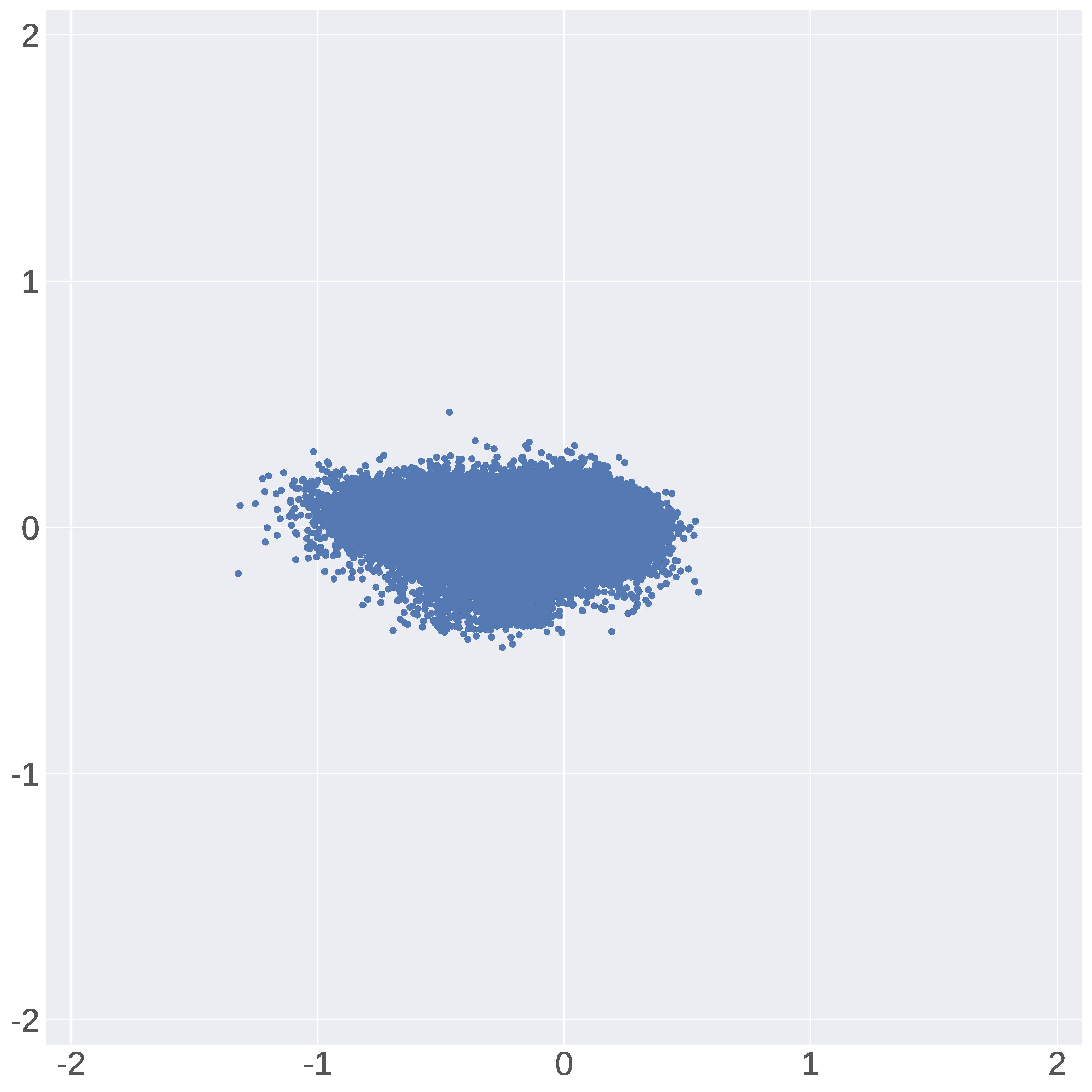} 
        \end{minipage}}%
    \subfloat[$\text{MCCLK}_{w/o \ L}$]
    {   \begin{minipage}[t]{0.19\linewidth}
            \centering     
            \includegraphics[width=\textwidth]{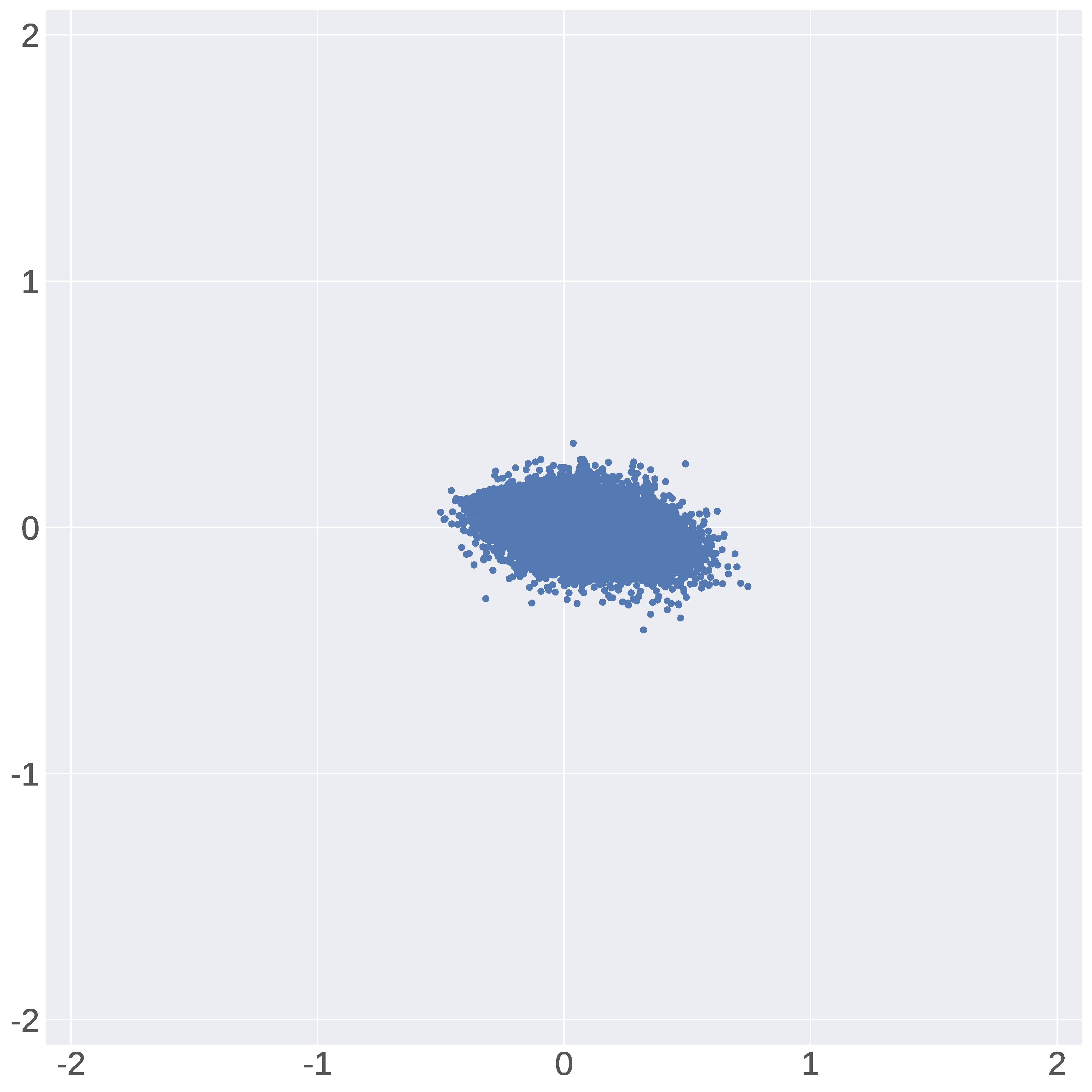} 
        \end{minipage}}%
    \subfloat[KGIN]
    {   \begin{minipage}[t]{0.19\linewidth}
            \centering     
            \includegraphics[width=\textwidth]{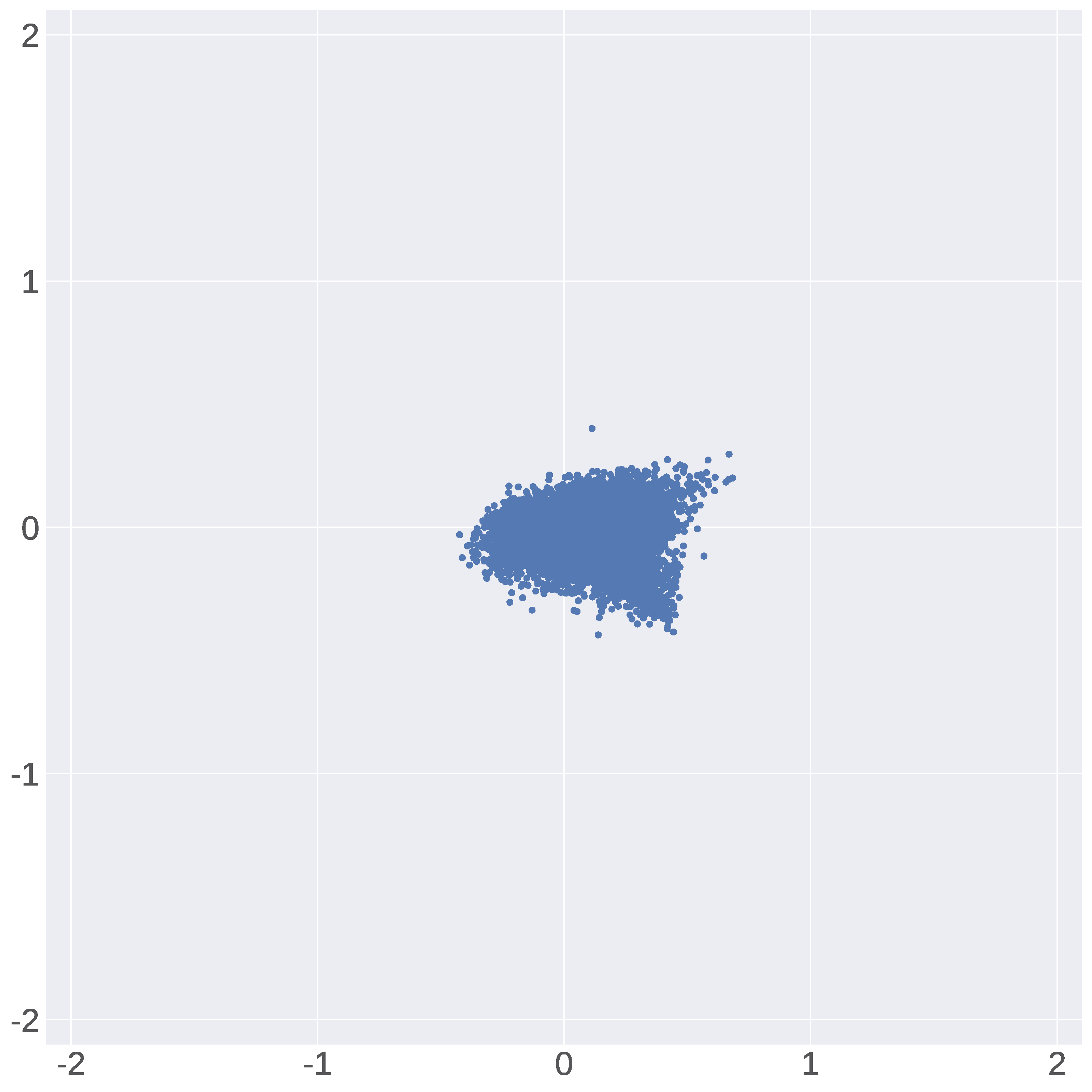} 
        \end{minipage}}%
    \subfloat[RippleNet]
    {   \begin{minipage}[t]{0.19\linewidth}
            \centering      
            \includegraphics[width=\textwidth]{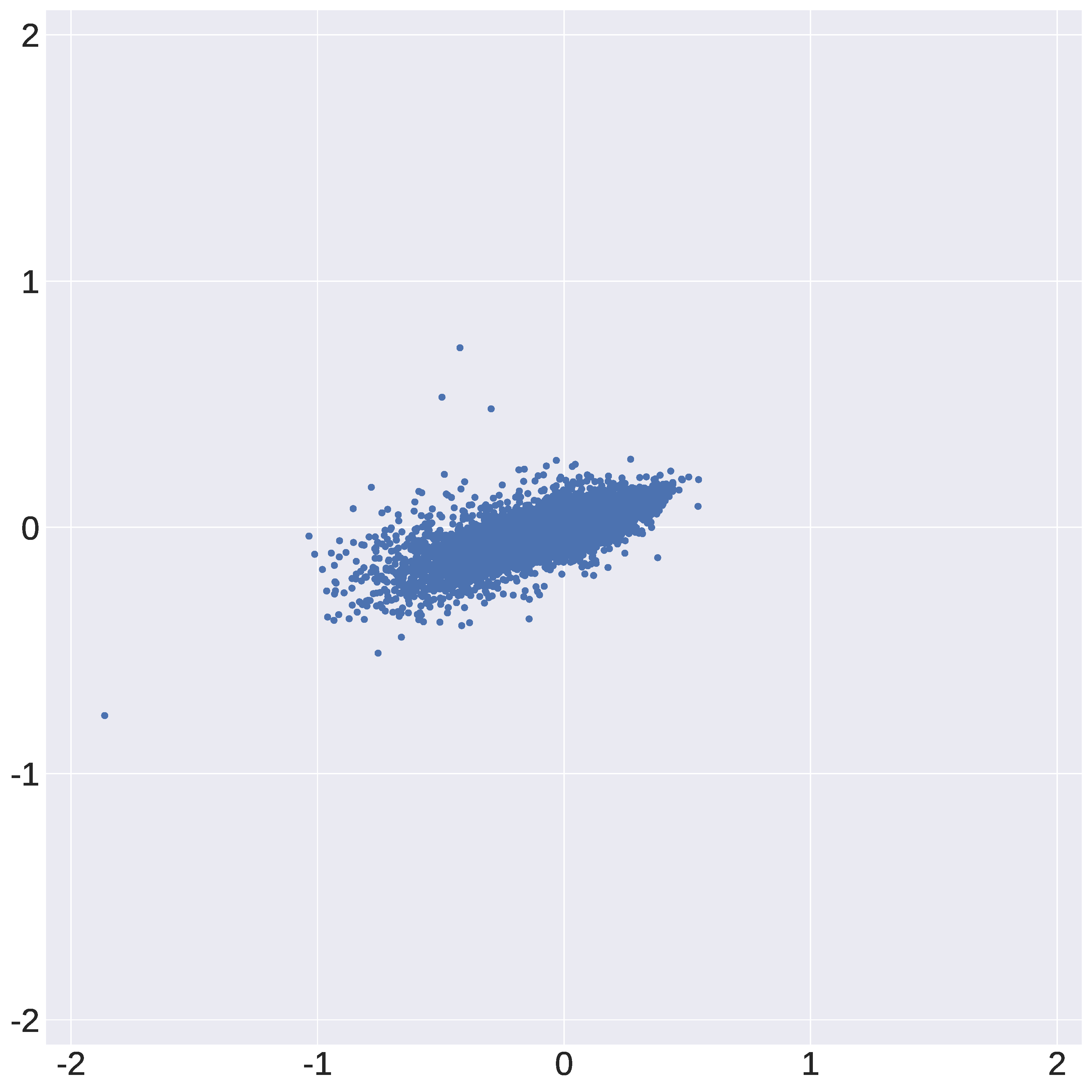}
        \end{minipage}}%
    \vspace{-0.4cm}
    \caption{Visualization of model representation learning ability on Book-Crossing.} 
    \label{fig::visual} 
\end{figure*}

\subsubsection{Impact of aggregation depth in semantic view.}

\begin{table}[t]
\centering
\setlength{\tabcolsep}{3pt}{
\begin{tabular}{l|c c|c c|c c}
\hline
 & \multicolumn{2}{c|}{Book} & \multicolumn{2}{c|}{Movie} & \multicolumn{2}{c}{Music} \\ 
 & Auc & F1 & Auc & F1 & Auc & F1 \\ \hline\hline
$L$=1 & \textbf{0.7602}	&\textbf{0.6777} & \textbf{0.9350}	&\textbf{0.8631} & 0.8711	&0.7858 \\ 
$L$=2 & 0.7601	&0.6768 & 0.9347	&0.8628 & \textbf{0.8742}	&\textbf{0.7945} \\ 
$L$=3 & 0.7591	&0.6733 & 0.9345	&0.8627 & 0.8726	&0.7891 \\ 
$L$=4 & 0.7583	&0.6749 & 0.9343	&0.8627 & 0.8720	&0.7846 \\ 
\hline
\end{tabular}}
\caption{Impact of aggregation depth in semantic view.}
\label{tab::itemitem}

\end{table}
To study the influence of item-item semantic graph aggregation depth, we vary $L$ in range of \{1, 2, 3, 4\} and demonstrate the performance comparison on book, movie, and music datasets in Table~\ref{tab::itemitem}. MCCLK performs best when $L=1, 1, 2$, on Book, Movie, and Music respectively. We can convince that: one- or two-hops are enough for aggregating neighbor information in the item-item semantic graph, which conveys the effectiveness of item-item semantic graph construction.

\subsubsection{Impact of aggregation depth in structural view}

\begin{table}[t]
\centering
\setlength{\tabcolsep}{3pt}{
\begin{tabular}{l|c c|c c|c c}
\hline
 & \multicolumn{2}{c|}{Book} & \multicolumn{2}{c|}{Movie} & \multicolumn{2}{c}{Music} \\ 
 & Auc & F1 & Auc & F1 & Auc & F1 \\ \hline\hline
$L'$=1 & 0.7602	&0.6776 & 0.9350	&0.8628 & 0.8711	&0.7858 \\
$L'$=2 & \textbf{0.7625}	&\textbf{0.6777} & \textbf{0.9351}	&\textbf{0.8631 }& \textbf{0.8763}	&\textbf{0.8008} \\ 
$L'$=3 & 0.7550	&0.6719 & 0.9334	&0.8589 & 0.8713	&0.7899 \\ 
$L'$=4 & 0.7569	&0.6680 & 0.9320	&0.8574 & 0.8706	&0.7841 \\ 
\hline
\end{tabular}}
\caption{Impact of aggregation depth in structural view.}
\label{tab::global}
\end{table}
To analyze the influence of aggregation depth in structural view, we vary $L'$ in range of \{1, 2, 3, 4\}, and illustrate the performance changing curves on book, movie, and music datasets in table~\ref{tab::global}. We find that: $L'=2$ are proper distance for collecting global structural signals from the longer-range connectivity (\ie u-r-i-r-e, i-r-e-r-i, \etc), further stacking more layers only introduces more noise.

\subsubsection{Impact of local-level contrastive loss weight $\alpha$.}

\begin{figure}[t] 
    \centering  
    \subfloat[Book]
    {   \begin{minipage}[t]{0.34\linewidth}
            \centering          
            \includegraphics[width=\textwidth]{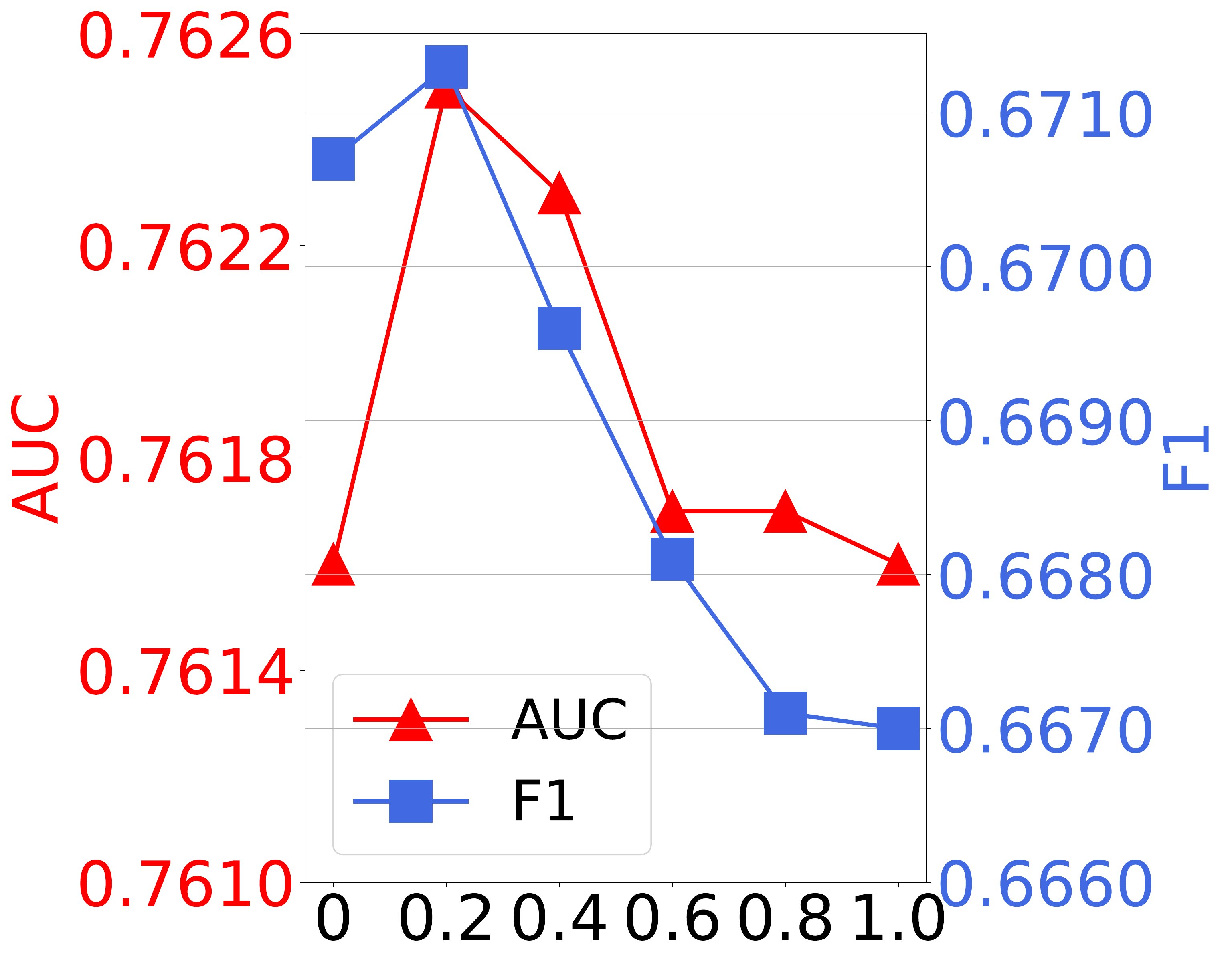} 
        \end{minipage}}
    \subfloat[Movie]
    {   \begin{minipage}[t]{0.34\linewidth}
            \centering     
            \includegraphics[width=\textwidth]{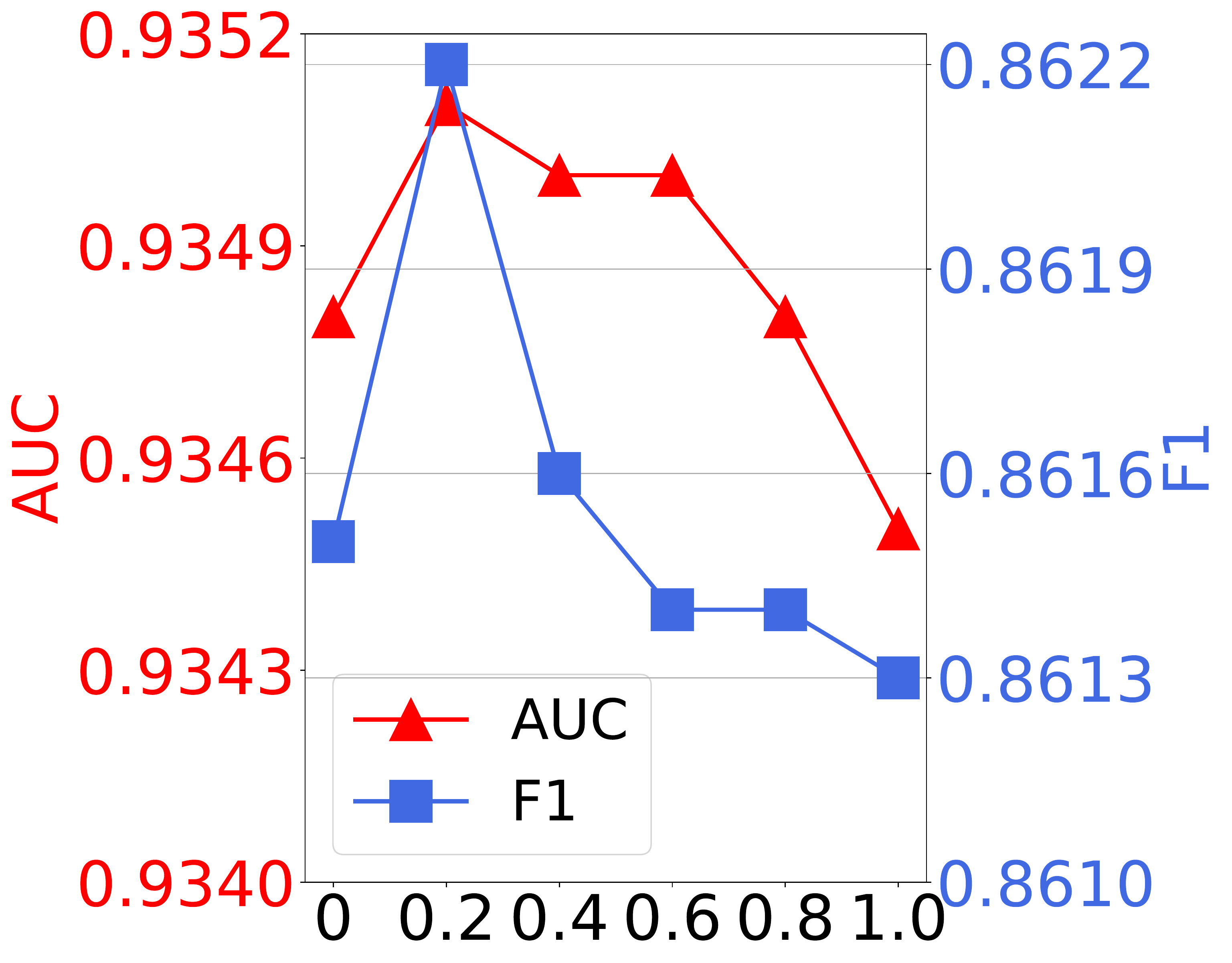} 
        \end{minipage}}%
    \subfloat[Music]
    {   \begin{minipage}[t]{0.34\linewidth}
            \centering      
            \includegraphics[width=\textwidth]{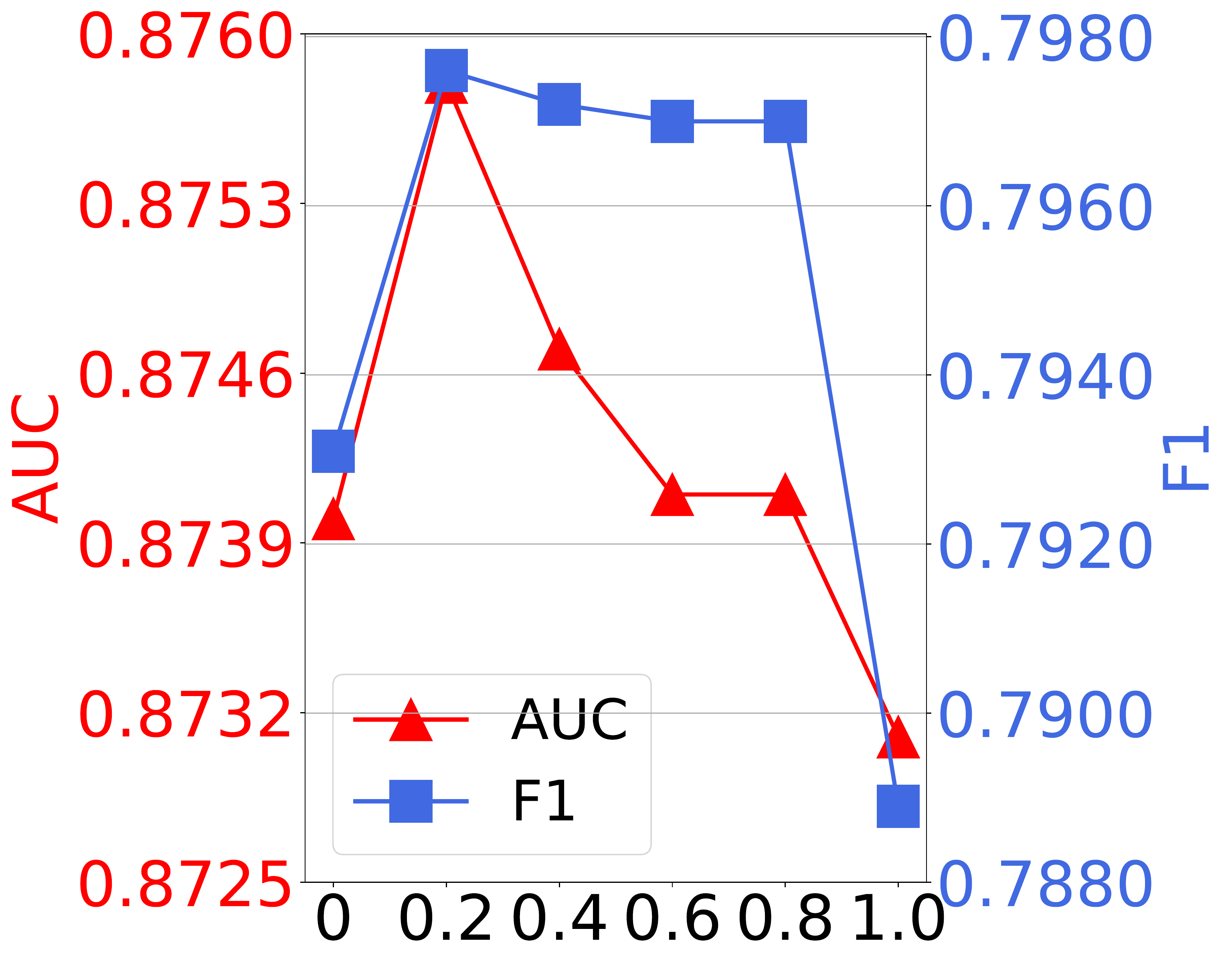}
        \end{minipage}}%
    \vspace{-0.4cm}
    \caption{Impact of local-level contrastive loss weight $\alpha$.} 
    \label{fig::alpha} 
\end{figure}
The trade-off parameter $\alpha$  controls the influence of local-level contrastive loss in final contrastive loss. To study the influence of $\alpha$, we vary $\alpha$ in \{0, 0.2, 0.4, 0.6, 0.8, 1.0 \}. According to the results shown in Figure~\ref{fig::alpha}, we have the following observations:
(1) The worst performance usually occurs when $\alpha=1$, which emphasizes the importance of global-level contrastive loss.
(2) The worse performance of scenarios where $\alpha=0, 1$ demonstrates the effectiveness of both two-level contrastive loss, and $\alpha=0.2$ balances local- and global-level contrastive loss on model optimization.

\subsubsection{Impact of contrastive loss weight $\beta$.}

\begin{table}[t]
\centering
\setlength{\tabcolsep}{3pt}{
\begin{tabular}{l|c c|c c|c c}
\hline
 & \multicolumn{2}{c|}{Book} & \multicolumn{2}{c|}{Movie} & \multicolumn{2}{c}{Music} \\ 
 & Auc & F1 & Auc & F1 & Auc & F1 \\ \hline\hline
$\beta$=1     & 0.7520 &0.6649 & 0.9337	&0.8593 & 0.8735 &0.7938 \\
$\beta$=0.1   & \textbf{0.7625} &\textbf{0.6713} & \textbf{0.9351} &\textbf{0.8622} & \textbf{0.8758} &\textbf{0.7972} \\ 
$\beta$=0.01  & 0.7608 &0.6689 & 0.9346 &0.8610 & 0.8721 &0.7913 \\
$\beta$=0.001 & 0.7607 &0.6675 & 0.9343	&0.8604 & 0.8714 &0.7856 \\ 
\hline
\end{tabular}}
\caption{Impact of contrastive loss weight $\beta$.}
\label{tab::beta}
\end{table}
The parameter $\beta$ determines the importance of the contrastive loss during the multi-task training. Towards studying the influence of contrastive loss weight $\beta$, we vary $\beta$ in \{1, 0.1, 0.01, 0.001\}
From the results shown in Table~\ref{tab::beta}, we can observe that: $\beta=0.1$ brings the best model performance, the main reason is that changing the contrastive loss to a fairly equal level to recommendation task loss could boast the model performance.
\subsection{Visualization (RQ4)}

To evaluate whether the contrastive mechanism affects the representation learning performance, following previous contrastive learning work \cite{qiu2021contrastive}, we adopt SVD decomposition to project the learned item embeddings into 2D and give out the regularized singular. As shown in Figure \ref{fig::visual}, we compare the visualized results of MCCLK, $\text{MCCLK}_{w/o \ G}$, $\text{MCCLK}_{w/o \ L}$, KGIN, and RippleNet on Book-Crossing, from which we can have the following observations:
\begin{itemize}
    \item The node embeddings of KGIN and RippleNet are mixed to some degree and fall into a narrow cone. In contrast, the node embeddings of MCCLK have a more diverse distribution and hence are able to represent different node feature information, which demonstrates our superiority in better representation learning and alleviating the representation degeneration problem.
    \item By comparing MCCLK with its variants, we observe that removing the local-level or global-level contrastive loss makes the learned embeddings more indistinguishable, which convinces the effectiveness and robustness of representation learning are coming from the multi-level cross-view contrastive learning mechanism. 
\end{itemize}

\section{Conclusion}

In this work, we focus on exploring contrastive learning on KG-aware recommendation, improving the quality of user/item representation learning in a self-supervised manner. We propose a novel framework, MCCLK, which achieves better user/item representation learning from two dimensions:
(1) MCCLK considers user/item representation learning from three views, including global-level structural view, local-level collaborative and semantic view, and explicitly construct a $k$NN item-item semantic graph to mine rarely noticed item-item semantic similarity in semantic view.
(2) MCCLK performs multi-level cross-view contrastive learning among three views, exploring both feature and structural information, and further learning discriminative representations.

\begin{acks}
    This work was supported in part by the National Natural Science Foundation of China under Grant No.61602197, Grant No.L1924068, Grant No.61772076, in part by CCF-AFSG Research Fund under Grant No.RF20210005, in part by the fund of Joint Laboratory of HUST and Pingan Property \& Casualty Research (HPL), and in part by the National Research Foundation (NRF) of Singapore under its AI Singapore Programme (AISG Award No: AISG-GC-2019-003). Any opinions, findings and conclusions or recommendations expressed in this material are those of the authors and do not reflect the views of National Research Foundation, Singapore. The authors would also like to thank the anonymous reviewers for their comments on improving the quality of this paper.
\end{acks}

\bibliographystyle{ACM-Reference-Format}
\bibliography{sample-base}

\end{document}